\documentclass[preprint,aps,showpacs,preprintnumbers,amsmath,amssymb,nofootinbib]{revtex4-1}
\pdfoutput=1
\usepackage{booktabs}
\usepackage{mathrsfs}
\usepackage{epsfig}
\usepackage{graphicx}
\usepackage{dcolumn}
\usepackage{bm}
\usepackage{amsmath}
\usepackage{slashed}       
\usepackage{amssymb}
\usepackage{amsfonts}
\usepackage{multirow}
\usepackage{makecell}
\usepackage{enumerate}
\usepackage[marginal]{footmisc}
\usepackage{float}
\usepackage{setspace}

\usepackage{color,xcolor,fancybox,epsf,rotating,colordvi}
\usepackage{ulem}
\def\add#1{\textcolor{black}{#1}}

\def\red#1{\textcolor{red}{#1}}

\def\blue#1{\textcolor{blue}{#1}}

\def\green#1{\textcolor{green}{#1}}

\def\gray#1{\textcolor{gray}{#1}}

\def\purple#1{\textcolor{purple}{#1}}

\def\addii#1{\textcolor{black}{#1}}

\let\jnfont=\rm
\def\NPB#1,{{\jnfont Nucl.\ Phys.\ B }{\bf #1},}
\def\PLB#1,{{\jnfont Phys.\ Lett.\ B }{\bf #1},}
\def\EPJC#1,{{\jnfont Eur.\ Phys.\ Jour.\ C }{\bf #1},}
\def\PRD#1,{{\jnfont Phys.\ Rev.\ D }{\bf #1},}
\def\PRL#1,{{\jnfont Phys.\ Rev.\ Lett.\ }{\bf #1},}
\def\MPLA#1,{{\jnfont Mod.\ Phys.\ Lett.\ A }{\bf #1},}
\def\JPG#1,{{\jnfont J.\ Phys.\ G}{\bf #1},}
\def\CTP#1,{{\jnfont Commun.\ Theor.\ Phys.\ }{\bf #1},}
\def\ZPC#1,{{\jnfont Z.\ Phys.\ C }{\bf #1},}
\def\JHEP#1,{{\jnfont JHEP \ }{\bf #1},}
\def\lsim{\raise0.3ex\hbox{$<$\kern-0.75em\raise-1.1ex\hbox{$\sim$}}}
\def\gsim{\raise0.3ex\hbox{$>$\kern-0.75em\raise-1.1ex\hbox{$\sim$}}}

\def\be{\begin{equation}}
\def\ee{\end{equation}}
\def\bea{\begin{array}}
\def\eea{\end{array}}
\def\beqa{\begin{eqnarray}}
\def\eeqa{\end{eqnarray}}
\def\beqas{\begin{eqnarray*}}
\def\eeqas{\end{eqnarray*}}

\def\bp{\begin{picture}}
\def\ep{\end{picture}}
\def\bc{\begin{center}}
\def\ec{\end{center}}
\def\bfig{\begin{figure}}
\def\efig{\end{figure}}

\def\bit{\begin{itemize}}
\def\eit{\end{itemize}}
\def\nn{\nonumber}
\def\f{\frac}

\def\[{\left[}
\def\]{\right]}
\def\({\left(}
\def\){\right)}

\def\..{\left.}
\def\.{\right.}
\def\tl{\tilde}

\def\tm{\times}

\def\zro{\bf 0}

\def\ep{\epsilon}

\newcommand{\GeV}{~{\rm GeV}}
\newcommand{\TeV}{~\rm TeV}

\begin{document}
\preprint{\parbox{1.2in}{\noindent arXiv: 1808.10851}}

\title{Solving the muon g-2 anomaly \add{in CMSSM extension} with non-universal gaugino masses 
}

\author{Fei Wang$^2$, Kun Wang$^1$, Jin Min Yang$^{3,4,5}$, Jingya Zhu$^{1,6}$ }

\address{
$^1$ Center for Theoretical Physics, School of Physics and Technology, Wuhan University, Wuhan 430072, P. R. China \\
$^2$ School of Physics, Zhengzhou University, ZhengZhou 450000, P. R. China \\
$^3$ CAS Key Laboratory of Theoretical Physics,
     Institute of Theoretical Physics,  Chinese Academy of Sciences, Beijing 100190,  P. R. China\\
$^4$ School of Physical Sciences, University of Chinese Academy of Sciences,
     Beijing 100049,  P. R. China\\
$^5$ Department of Physics, Tohoku University, Sendai 980-8578, Japan\\
$^6$ Enrico Fermi Institute, University of Chicago, Chicago, IL 60637, USA
}

\begin{abstract}
We propose to generate non-universal gaugino masses in SU(5) Grand Unified Theory (GUT) with the generalized Planck-scale mediation
SUSY breaking mechanism, in which the non-universality arises from proper wavefunction normalization with lowest component
VEVs of various high dimensional representations of the Higgs fields of SU(5) and an unique F-term VEV by the singlet.
Different predictions on gaugino mass ratios with respect to widely studied scenarios are given.
The gluino-SUGRA-like scenario, where gluinos are much heavier than winos, bino and universal scalar masses,
can be easily realized with appropriate combinations of such high-representation Higgs fields. With six GUT-scale
free parameters in our scenario, we can solve 
\add{elegantly} the tension between mSUGRA and the present experimental
results, including the muon g-2, the dark matter (DM) relic density and the direct sparticle search bounds from the LHC.
Taking into account the current constraints in our numerical scan, we have the following observations:
\add{(i) The large-$\tan\beta$ ($\gtrsim35$) samples with a moderate $M_3$ ($\thicksim 5\TeV$), a small $|A_0/M_3|$ ($\lesssim0.35$) and a small $m_A$ ($\lesssim 4\TeV$) are favoured to generate a 125 GeV SM-like Higgs
and predict a large muon g-2, while the stop mass and $\mu$ parameter, mainly determined
by $|M_3|$ ($\gg M_0,|M_1|,|M_2|$), can be about 6 TeV;}
\add{(ii) The moderate-$\tan\beta$ ($35\thicksim40$) samples with a negative $M_3$ can have a
light smuon ($250\thicksim450$ GeV) but a heavy stau ($\gtrsim 1\TeV$), which predict a large muon g-2
but a small $Br(B_s\to\mu^+\mu^-)$;}
(iii) To obtain the right DM relic density, the annihilation mechanisms should be stau exchange, stau coannihilation,
chargino coannihilation, slepton annihilation and the combination of two or three of them;
(iv) To obtain the right DM relic density, the spin-independent DM-nucleon cross section is typically
much smaller than the present limits of XENON1T 2018
and also an order of magnitude lower than the future detection sensitivity of LZ and XENONnT experiments.
\end{abstract}


\maketitle

\section{Introduction}
Low energy supersymmetry (SUSY), which is well motivated to solve the hierarchy problem, is one of the most appealing
new physics candidates beyond the Standard Model (SM).
The gauge coupling unification, which cannot be realized in the SM, can be successfully realized in the framework
of weak scale SUSY.
Besides, assuming R-parity conservation, the lightest SUSY particle (LSP) can be a promising dark matter (DM) candidate
with the right DM relic density.

However, there are over 100 physical free parameters in the minimal SUSY model (MSSM),
including the soft masses, phases and mixing angles that cannot be rotated away by redefining the phases
and flavor basis for the quark and lepton supermultiplets.
In practice, some universalities of certain soft SUSY breaking parameters as high scale inputs are usually adopted.
In the constrained MSSM (CMSSM), the gaugino masses, the sfermion masses and the trilinear couplings
are all assumed to be universal at the GUT scale, respectively.
Thus, CMSSM only has five free parameters, i.e., $\tan\beta,\,M_0,\,A_0,\,M_{1/2}$ and the sign of $\mu$.
All the low energy soft SUSY breaking parameters can be determined by these five inputs through the
renormalization group equations (RGEs) running from the GUT scale to the EW scale.

So far the null search results of the gluino and the first two generations of squarks
together with the 125 GeV Higgs discovery \cite{1207-a-com,1207-c-com} at the LHC
have severely constrained the parameter space of CMSSM.
For example, to provide a 125 GeV SM-like Higgs, the stop masses should be around 10 TeV or the trilinear stop
mixing parameter $A_t$ should be quite large.
Besides, in order for the gluino to escape the LHC bounds, the universal gaugino mass at the GUT scale $|M_{1/2}|$
should be larger than about 1 TeV ($m_{\tl{g}}\simeq 2|M_{1/2}|$ ),
and thus the bino-like neutralino is bounded to be higher than about 400 GeV.
All the electroweakinos (higgsinos, sleptons, sneutrinos) are all bounded to be heavier than several hundreds
of GeV in CMSSM, and hence
the SUSY contributions to the muon g-2 cannot be large enough to account for the discrepancy reported by
the Brookhaven AGS \cite{mug2-ex2006, cmssm-125, cmssm-zy}.
\add{In fact, CMSSM was found to be excluded at $90\%$ confidence level \cite{cmssm-killing} if it is required to account for both the muon g-2 anomaly and the recent LHC constraints on SUSY particles.}
The neutralino dark matter in CMSSM, which is heavier than several hundreds of GeV,
can mainly have four annihilation mechanisms: stau coannihilation, stop coannihilation, A/H funnel, hybrid
(note that the h/Z funnel cannot happen and the focus point or $\chi^{\pm}_1$ coannihilation is severely constrained by the recent dark matter direct detection limits) \cite{susy-dm, cmssm-zy}.
\add{We should mention that even if only the DM relic density upper bound is considered in addition to the muon g-2, a global fit by the GAMBIT \cite{gambit:CMSSM} collaboration still disfavors the CMSSM.}

Motivated by the tension between CMSSM and the experiment results, several extensions of CMSSM have been proposed,
such as introducing right-handed neutrinos \cite{cmssm-neu}, singlet scalars \cite{cnmssm} or vector-like supermultiplets \cite{cmssm-vlf}.
Other proposals were also considered, such as relaxing universal conditions at the GUT scale \cite{cmssm-nuhm, cmssm-nugm, nugm-Martin} or reducing the universal conditions
to certain sub-GUT scale \cite{cmssm-subgut} (such as the `mirage' unification scale \cite{FW-mirage2018})
or including the reheating temperature in the early universe as an extra parameter \cite{cmssm-eu}.
Among these approaches, the non-universal gaugino mass (NUGM) \cite{nugm-first} scenarios are well motivated
both theoretically and phenomenologically,
which can be realized by some special structure of gauge kinetic function in string models \cite{nugm-string}
or the GUT \cite{nugm-gut}.

\add{In this work, we propose a NUGM extension of mSUGRA, where the non-universality arises from proper wavefunction normalization with lowest component VEVs of various high dimensional representations of the Higgs fields of SU(5) and an unique F-term VEV by the singlet.}
By properly choosing the parameters,
we can generate the gaugino hierarchy $M_3\gg M_1,M_2$ and obtain a low energy spectrum which can
escape the LHC mass bounds for colored sparticles and at the same time give the 125 GeV Higgs mass.
The muon g-2 anomaly can also be solved with $M_0$ at a few hundreds of GeV.
Besides, the flavour bounds, the LHC direct search bounds as well as the updated dark matter constraints
can all be satisfied. Note that this setting can be fitted into the $\tl{g}\rm SUGRA$ scenario proposed in \cite{nath}.

This paper is organized as follows. We propose our approach to generate non-universal gaugino masses in SU(5) GUT with the generalized Planck-scale mediation of SUSY breaking in Section II.
In Section III, we discuss the constraints on the model.
In Section IV, we present our numerical results and give some discussions.
Finally, we draw our conclusions in Section V.

\section{Generating non-universal gaugino masses in SU(5)}
As a low energy phenomenological model, the MSSM is not very predictive because it has more than
one hundred free parameters.
The gaugino mass hierarchy, which is also not specified in MSSM, is very important in understanding the nature
of dark matter.
For example, possible signals for neutralino dark matter direct detection experiments depend on the ratio of
each component of the dark matter particle. So it is desirable to seek for some UV-completed models that can
predict the low energy soft SUSY breaking parameters with few input parameters.

Some UV-completed models of MSSM, e.g., the gravity mediated SUSY breaking scenarios with the simplest Kahler potential,
predict universal gaugino masses at the GUT scale. After RGE running to EW scale, the approximate ratio for gaugino masses\footnote{Such a ratio at the EW scale is also predicted by the minimal gauge mediated SUSY breaking mechanism.}
is $M_1:M_2:M_3\approx 1:2:6$.
We know that the latest LHC results have pushed the gluino mass to about 2 TeV, and thus the neutral electroweakinos
are also heavy and cannot solve the muon $g-2$ anomaly.
Actually, the gaugino masses at the GUT scale are not necessarily universal.
In realistic SUSY GUT models, certain high dimensional GUT group representations of Higgs fields
may play an essential role in solving the doublet-triplet splitting problem or generating realistic
fermion ratios if Yukawa unification is further assumed. With such high dimensional Higgs fields,
the universal soft SUSY breaking masses can receive additional non-universal parts.
\add{For example, the scenarios with non-universal gaugino masses have been studied
in \cite{nugm1, nugm2, nugm-mug2, nugm-dm, King:2007vh} and references therein.}
General results of soft SUSY breaking parameters in the generalized SUGRA \cite{LN, BLNW} for SU(5), SO(10) and E6 GUT models involving various high dimensional Higgs fields
with different symmetry breaking chains have been discussed in \cite{BLNW,fei}.
Some applications have been also studied \cite{tjl1,FWY2}.

The gaugino masses can always be given by the following non-renomalizable superpotential terms
\beqa
W\supseteq \f{f_a}{4}\[ W^aW^a+ a_1\f{T}{\Lambda} W^a W^a+ b_1\f{1}{\Lambda} W^a \Phi_{ab} W^b+c_1\f{T}{\Lambda^2} W^a \Phi_{ab} W^b \],
\label{nugm-W}
\eeqa
with $\Lambda$ being a typical UV energy scale (say the Planck scale $M_{Pl}$) upon the GUT scale.
The chiral superfield $T$ is a GUT group singlet and $\Phi_{ab}$ is a chiral superfield lying in any of the
irreducible representations within the symmetric group production decomposition of adjoint representations.
For example, in the framework of SU(5) GUT, $\Phi_{ab}$ can belong to
\begin{eqnarray}
  \bf (24\otimes 24)_{\rm symm} = 1 \oplus 24 \oplus 75 \oplus 200.
\end{eqnarray}
After $\Phi_{ab}$ or $T$ acquiring an F-term VEVs, soft SUSY breaking gaugino masses will be predicted.
For example, the term proportional to $a_1$ will generate universal gaugino masses with
non-vanishing $\langle F_T\rangle$, while the term proportional
to $b_1$ will generate non-universal gaugino parts with non-vanishing $\langle F_{\Phi_{ab}}\rangle$.
\add{In most of the previous studies, non-vanishing F-term VEVs of the GUT non-singlet Higgs field $\Phi_{ab}$
are necessarily present to generate non-universal gaugino masses.}
In principle, the soft sfermion masses or trilinear couplings may also receive additional non-universal
contributions from such high dimensional operators.

\add{Although it is indeed possible to realize SUSY breaking with a F-term VEV for a gauge non-singlet superfield
through model buildings, for example in typical dynamical SUSY breaking models or ISS-type models,
it is more natural to realize SUSY breaking with a gauge singlet F-term VEV.}
We propose a new approach in which the leading non-universality of gaugino masses comes from
the wavefunction normalization \add{with a F-term VEV for a gauge singlet only} \footnote{\add{We should note that it is possible to generate subleading non-universal gaugino masses from the $c_1$ term of Eq.(\ref{nugm-W}) with a
suppression factor $\Lambda^{-2}$ in comparsion with the leading universal gaugino mass part which is
suppressed by $\Lambda^{-1}$.}}.
\add{Although simple, this possibility, which leads to different predictions on gaugino hierarchy,
has not been emphasized and discussed in previous non-universal gaugino masses literatures.}

We can consider the most general combination involving the ${\bf 24}, {\bf 75}$ and ${\bf 200}$ representations
of Higgs fields of SU(5) GUT group and the gauge singlet $T$
\beqa
{\cal L}=\int d^2\theta \(\f{f_a}{4} W^a W^b\) \f{1}{\Lambda}\[\f{}{}c_0 T \delta_{ab} + c_1 (H_{\bf 24})_{ab}+c_2 (H_{\bf 75})_{ab}+c_3 ( H_{\bf 200})_{ab}\]~.
\eeqa
\add{In previous studies on non-universal gaugino masses, as noted previously, non-vanishing $F_{H_r}$
are necessarily present to generate non-universal gaugino masses with (almost) trivial kinetic terms for gauginos.}
It is however possible that only the GUT group singlet $T$ acquires both F-term VEV $\langle F_T\rangle$ and lowest component VEV $\langle T\rangle$
while all other high dimensional representation Higgs fields acquire only the lowest component VEVs
that still preserve the $SU(3)_C\tm SU(2)_L\tm U(1)_Y$ gauge group
\beqa
\langle T\rangle= T_0+\theta^2 F_T~,~~~~\langle H^r_{ab}\rangle=v_U M^r_{ab}
\eeqa
with $M^r_{ab}$ being the group factor for the representation $r$ and $v_U$ the GUT breaking scale
which is assumed to be independent of the SU(5) representation and universal for all $H^r_{ab}$.
The VEVs of the Higgs field $\Phi_{\bf 24}$ in the adjoint representation can be expressed as a $5\times 5$
matrix
\beqa
\langle \Phi_{\bf 24} \rangle ~=~ v_U {\sqrt {3\over 5}} {\rm diag}
\left(-{1\over 3}, -{1\over 3}, -{1\over 3},
{1\over 2}, {1\over 2} \right)~,~\,
\eeqa
while the VEVs of the Higgs field $\Phi_{\bf 75}$ can be expressed as a $10\times 10$
matrix
\beqa
\langle \Phi_{\bf 75} \rangle ~=~\f{v_U}{2\sqrt{3}}
{\rm diag}
\left(\underbrace{~1,\cdots,~1}_3,\underbrace{-1,\cdots,-1}_{6},
3\right)~.
\eeqa
Similarly, the VEVs of the Higgs field $\Phi_{\bf 200}$ can be expressed as a $15\times 15$ matrix
\beqa
\langle \Phi_{\bf 200} \rangle ~=~ \f{v_U}{2\sqrt{3}} {\rm diag}
\left(\underbrace{~1,\cdots,~1}_6,\underbrace{-2,\cdots,-2}_{6},\underbrace{~2,\cdots,~2}_{3}
\right)~.~ \,
\eeqa
As $T_0$ is a GUT group singlet, the VEV $T_0$ can be of order $\Lambda$ without spoiling GUT.
The kinetic term after substituting the lowest component VEV will take the form
\beqa
W\supseteq \f{f_a}{4}W^aW^b\[ (1+ a_1\f{T}{\Lambda})\delta_{ab} + \sum\limits_{r}c_i\f{v_U}{\Lambda} \langle M^r_{ab}\rangle \].
\eeqa
As $v_U\ll \Lambda$ and $T_0\simeq \Lambda$, the term proportional to $\delta_{ab}$ will be the leading normalization factor.
If this term nearly vanishes by choosing a proper $a_1$, the second term, which is non-universal for three gauge couplings,
will generate a different wavefunction normalization factor. On the other hand, substituting the F-term VEV $F_T$ will
generate universal gaugino masses for non-canonical gauge fields.
After normalizing the gauge kinetic term to canonical form, non-universal gaugino masses will be generated.
The prediction in this scenario is different from previous studies.
\add{In Table \ref{ratios}, we list our prediction for gaugino mass ratios in different SU(5) representations, in comparison with previous studies (e.g., Ref.\cite{King:2007vh}).}
For example, if only the ${\bf 24}$ representation Higgs is present other than $T$,
the gaugino ratio is given by
\beqa
M_1:M_2:M_3=1:\f{1}{3}:-\f{1}{2}~,
\eeqa
at the GUT scale which will predict the gaugino ratio $3:2:-9$ at the EW scale.
\add{So the wino is the lightest gaugino and possibly be the DM candidate in contrary to the widely studied scenarios with bino as the lightest gaugino for $\langle F_{\bf 24} \rangle\neq 0$.}
\add{Another example, although we adopt the most general form of combinations, gluino SUGRA can in fact be realized with only one 200 or 75 representation, in which the gluino can be much (almost 5$\sim$10 times) heavier than wino and bino at EW scale.}
For the most general combinations involving all ${\bf 24}, {\bf 75}$
and ${\bf 200}$ Higgs fields, we can obtain the gaugino ratio $M_1:M_2:M_3$ as
\beqa
\label{nug1}
\[-\f{c_1}{4\sqrt{15}}+\f{5c_2}{4\sqrt{3}}+\f{5c_3}{2\sqrt{3}}\]^{-1}:
\[-\f{3c_1}{4\sqrt{15}}-\f{3c_2}{4\sqrt{3}}+\f{c_3}{2\sqrt{3}}\]^{-1}:\[\f{c_1}{2\sqrt{15}}-\f{c_2}{4\sqrt{3}}+\f{c_3}{4\sqrt{3}}\]^{-1},\nn\\
\eeqa
at the GUT scale. So we can see that one can get an arbitrary gaugino ratio at GUT scale with different choices of $c_i$.
This result is different from the widely studied scenarios in which both the high-representation Higgs fields $H_{ab}$ and $T$ acquire universal F-term VEVs $F_U$ with trivial kinetic terms, which gives the gaugino mass ratio $M_1:M_2:M_3$
at the GUT scale as
\beqa
\label{nug}
\[c_0-\f{c_1}{4\sqrt{15}}+\f{5c_2}{4\sqrt{3}}+\f{5c_3}{2\sqrt{3}}\]:
\[c_0-\f{3c_1}{4\sqrt{15}}-\f{3c_2}{4\sqrt{3}}+\f{c_3}{2\sqrt{3}}\]:\[c_0+\f{c_1}{2\sqrt{15}}-\f{c_2}{4\sqrt{3}}+\f{c_3}{4\sqrt{3}}\],\nn\\
\eeqa
\add{So, an arbitrary gaugino ratio of $M_1:M_2:M_3$ at the GUT scale can be obtained with properly chosen coefficients $c_1,c_2,c_3$ or $c_0,c_1,c_2,c_3$. Besides, the new and old approaches will in general lead to different predictions on the nature of lightest gaugino.}

\begin{table}[!htbp]
\vspace*{-0.3cm}
\centering
\caption{\add{Comparison of our predictions with previous studies (e.g., Ref.\cite{King:2007vh})
on gaugino mass ratios in different SU(5) representations, where `GUT' is for results at GUT scale
and `EW' is for results at EW scale.}}
\label{ratios}
\begin{spacing}{1.2}
\begin{tabular}{|c|c|c|}
\hline\hline
representations & gaugino ratios in our work & ~~gaugino ratios in previous studies~~
\\ \hline
{\bf 1} & GUT $1:1:1$, ~~EW $1:2:6$ & GUT $1:1:1$, ~~EW $1:2:6$
\\ \hline
{\bf 24} & GUT $1:\frac{1}{3}:-\frac{1}{2}$, ~~EW $3:2:-9$ & GUT $1:3:-2$, ~~EW $1:6:-12$
\\ \hline
{\bf 75} & ~~GUT $-\frac{1}{5}:\frac{1}{3}:1$, ~~EW $-3:10:90$~~ & GUT $-5:3:1$, ~~EW $-5:6:6$
\\ \hline
{\bf 200} & GUT $\frac{1}{10}:\frac{1}{2}:1$, ~~EW $1:10:60$ & GUT $10:2:1$, ~~EW $5:2:3$
\\ \hline\hline
\end{tabular}
\end{spacing}
\end{table}

An interesting region will appear if $M_3\gg M_2,M_1$. In this region, the colored sfermions are heavy even
for a very small $M_0$ (which is the universal sfermion mass parameter) because of the loop corrections involving a heavy $M_3$. The non-colored sfermions will,
however, still be light if the GUT scale mass $M_{1,2}\ll M_3$.
This region, which is called the gluino-SUGRA region \cite{nath}, is well motivated to solve the muon
$g-2$ anomaly \cite{FWY} and at the same time be consistent with the LHC predictions.
In our new approach, the gluino-SUGRA region is easily realized if the denominators of the third term within Eq.(\ref{nug1}) nearly vanishes
\beqa
\f{c_1}{2\sqrt{15}}-\f{c_2}{4\sqrt{3}}+\f{c_3}{4\sqrt{3}}\approx 0~,
\eeqa
while the denominators of the first two terms are non-vanishing.
In the widely studied approach which is given by Eq.(\ref{nug}), to realize the gluino-SUGRA region, the first two terms within the second line
of Eq.(\ref{nug}) need to vanish approximately while the third term should be non-vanishing.
Solving for $c_1,c_2$ in terms of $c_0$ and $c_3$ gives
\beqa
c_1&=&\f{20\sqrt{5}}{9}c_3+\f{16\sqrt{5}}{9}c_0~,~c_2=-\f{14}{9}c_3-\f{4\sqrt{3}}{9}c_0~,
\eeqa
where $c_3\neq -\f{8\sqrt{3}}{7}c_0$.

\section{The scan and constraints}

In order to illustrate the salient features of our scenarios, we scan the six dimensional parameter space considering all current experimental constraints. The package NMSSMTools \cite{nmssmtools} is used in our numerical scan to obtain the low energy SUSY spectrum.
We know that in case $\lambda\sim\kappa\to 0$ and $A_\lambda$ is small, the singlet superfield within the NMSSM
will decouple from other superfields and the NMSSM will reduce to the MSSM plus a decoupled heavy singlet scalar
and singlino. So the MSSM spectrum can be calculated with NMSSMTools.
In our scan, we use the program $\textsf{NMSPEC\_MCMC}$ \cite{nmspec} in
$\textsf{NMSSMTools\_5.2.0}$ \cite{nmssmtools}.
The ranges of parameters in our scan are
\begin{eqnarray}
\label{eq:space}
1{\TeV} <|M_3|<10 {\TeV}, \quad & \quad |A_0| < 20 {\TeV} ,
\nonumber\\
\quad M_0,|M_1|,|M_2|<1 {\TeV}, \quad & \quad 1<\tan\beta <60,
\end{eqnarray}
where we choose a large $|M_3|$ to escape the LHC bounds on colored sparticles
and a large $|A_0|$ to generate the 125 GeV Higgs mass.
Small $M_0, |M_1|, |M_2|$ and a large $\tan\beta$ are chosen to give large SUSY contributions
to the muon g-2  and a low mass for dark matter particle.

In our scan, we consider the following constraints:
\begin{enumerate}[(1)]
  \item The theoretical constraints of vacuum stability, and no Landau pole below $M_{\rm GUT}$ \cite{nmssmtools, nmspec}.
  \item The lightest CP-even Higgs boson $h$ as the SM-like Higgs boson with a mass of 123-127 GeV.
      \add{The SM-like Higgs mass is calculated including corrections of top/bottom sector at full 1 loop and leading logs 2 loop level, and other sfermions, charginos/neutralinos, and gauge bosons at leading logs 1 loop level.}
      \footnote{\add{With fixed-order method, full one-loop corrections have been calculated diagrammatically in Ref.\cite{MSSM-mh1L}, dominant two-loop corrections in Ref.\cite{MSSM-mh2L} and partial three-loop corrections in Ref.\cite{MSSM-mh3L}.
      Also, effective field theory (EFT) methods were applied in Ref.\cite{MSSM-mh-EFT}, and fourth logarithmic order corrections was calculated in Ref.\cite{MSSM-mh4L}.
      Besides, hybrid methods have also been developed \cite{MSSM-mh-hybrid}, and been implemented in the publicly available code \textsf{FeynHiggs} \cite{FeynHiggs}. We checked our Higgs masses with \textsf{FeynHiggs-2.14.3}, and found they coincide at 2 GeV level for most samples.}}
    Its production rates should fit the LHC data globally \cite{15007-Atlas, 1412-CMS} with the method in Refs.\cite{low-NMSSM,1311-mdm}.
  \item The searches for low mass and high mass resonances at the LEP, Tevatron, and LHC,
        which constrained the production rates of heavy Higgs bosons.
      We implement these constraints by the package \textsf{HiggsBounds-5.1.1beta} \cite{higgsbounds}.
  \item The constraints for squarks and gluino from the LHC:
        \begin{eqnarray}
        m_{\tilde{g}} \gtrsim 2 {\TeV}, \quad m_{\tilde{t}} \gtrsim 0.7 {\TeV},
        \quad m_{\tilde{q}_{1,2}} \gtrsim 2 {\TeV},
        \end{eqnarray}
        and the lower mass bounds of charginos, sleptons from the LEP:
        \begin{eqnarray}
        m_{\tilde{\tau}} \gtrsim 93.2 {\GeV}, \quad m_{\chi^{\pm}} \gtrsim 103.5 {\GeV}.
        \end{eqnarray}
  \item The searches for chargino $\chi^{\pm}_1$ and next-to-lightest neutralino $\chi^0_2$ at the LHC:
      \begin{eqnarray}
      pp\to \chi^+_1\chi^-_1, \; \chi^{\pm}_1\chi^0_2
      \end{eqnarray}
      For these searches, we only employ the channels with tau leptons in final states \cite{1708-a-ew}
      because we checked that the dominated decay channels of $\chi^{\pm}_1$ and $\chi^0_2$ are
      \begin{eqnarray}
      \chi^{\pm}_1 \to \tilde{\tau}^{\pm} \nu_{\tau} , ~~~&~~
      \chi^{0}_2 \to \tilde{\tau}^{\pm} \tau^{\mp}  ~~~&~~~(~ {\rm for}~ m_{\chi^{\pm}_1}>m_{\tilde{\tau}}),
      \nonumber\\
      \chi^{\pm}_1 \to \tau^{\pm}\nu_{\tau} \chi^0_1, &~~
      \chi^{0}_2 \to \tau^{\pm}\tau^{\mp} \chi^0_1 &~~~(~ {\rm for}~ m_{\chi^{\pm}_1}<m_{\tilde{\tau}}).
      \end{eqnarray}
  \item Constraints from B physics, such as $B \to X_s \gamma$, $B_s \to \mu^+ \mu^-$,  $B_d \to \mu^+ \mu^-$ and $B^+ \to \tau^+ \nu_\tau$, and the mass differences $\Delta M_d$, $\Delta M_s$ \cite{BaBar-Bph, LHCb-BsMuMu, PDG2016}
      \begin{eqnarray}
        1.7\times10^{-9} <&Br(B_s\to\mu^+\mu^-)&< 4.5\times10^{-9} , \nonumber\\
        1.1\times10^{-10} <&Br(B_d\to\mu^+\mu^-)&< 7.1\times10^{-10}, \nonumber\\
        2.99\times10^{-4} <&Br(b\to s\gamma)&< 3.87\times10^{-4}.
      \end{eqnarray}
  \item The constraints of muon anomalous magnetic moment (muon g-2) at 2$\sigma$ level.
For the experimental data and SM calculation without the Higgs contribution
(because there is a SM-like Higgs boson in SUSY models contributing to $\Delta a_\mu$), we use \cite{mug2-ex2006, mug2-sm}
      \begin{eqnarray}
        a_\mu^{\rm ex} &=& (11659208.0 \pm 6.3) \times 10^{-10}, \\
        \Delta a_\mu &\equiv& a_\mu^{\rm ex}-a_\mu^{\rm SM} = (27.4\pm9.3) \times 10^{-10}.
      \end{eqnarray}
      We calculate the SUSY contribution $\Delta a_\mu$ including the SM-like Higgs boson, and require $\Delta a_\mu$ to lie at the $2\sigma$ level.
      We also include the theoretical uncertainty in SUSY $\Delta a_\mu$ calculations, which is about
      \add{$\delta^{\rm th}\approx3 \times 10^{-10}$ \cite{mug2-susy06}}
      \footnote{\add{The $\Delta a_\mu$ includes chargino \cite{mug2-susy06}, neutralino \cite{mug2-neu}, and Higgs \cite{mug2-hig} contributions, all at 2-loop level.
      The theoretical uncertainty is calculated as $\delta^{\rm th}\equiv2.8\times 10^{-10}+0.02\left|\Delta a_\mu^{\rm 1L}\right| +0.3\left|a_\mu^{\rm 2L}-a_\mu^{\rm 1L}\right|$ \cite{mug2-susy06}, a little larger than that in Ref. \cite{mug2-therr}.
      And the theoretical uncertainty is added linearly to $\Delta a_\mu$, totally required to satisfy $a_\mu^{\rm ex}-a_\mu^{\rm SM}$ at $2\sigma$ level.}}.
  \item Constraints from dark matter relic density by WMAP/Planck \cite{dm-omg, PDG2016}, and the 2018 result of direct searches for dark matter at XENON1T \cite{XENON1T2018}.
      We require the lightest neutralino $\chi^0_1$ to be the dark matter candidate\add{ and calculate its relic density and cross sections by \sf{micrOMEGAs} \cite{micromegas} inside \sf{NMSSMTools}}.
      For DM relic density, we only apply the upper bound, e.g., $0\leq \Omega \leq 0.131$, as other dark matter species may also contribute to the DM relic density \cite{cmssm-eu, low-NMSSM98, nmssmtools}.
      \add{For DM-nucleon scattering cross section, we rescale the original values by $\Omega/\Omega_0$ with $\Omega_0 h^2=0.1187$ to impose the XENON1T constraint.}
\end{enumerate}

We take a multi-path Markov Chain Monte Carlo (MCMC) scan in the parameter space.
In total, we get nearly $10^7$ surviving samples.

\section{Numerical results and discussions}
\begin{figure}[htb]
  \centering
\includegraphics[width=16cm]{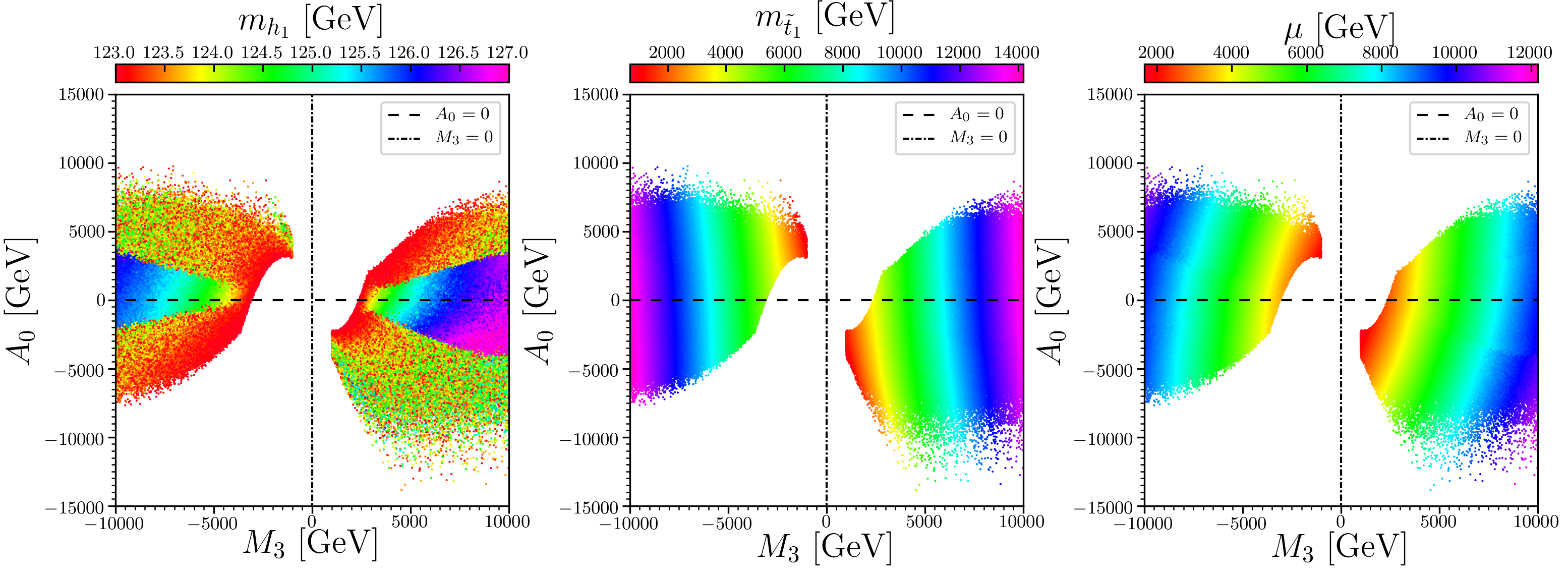}
\vspace*{-0.5cm}
\caption{Surviving samples in the $A_0$ versus $M_3$ planes, with colors indicating
the lighter Higgs mass $m_h$ (left), lighter stop mass $m_{\tilde{t}_1}$ (middle) and
higgsino mass parameter $\mu$ (right), respectively.
\add{In these 2-dimension planes, larger-$\Delta a_\mu$ samples are projected on top of smaller ones. }}
\label{fig1}
\end{figure}
\begin{figure}[!htb]
  \centering
\includegraphics[width=16cm]{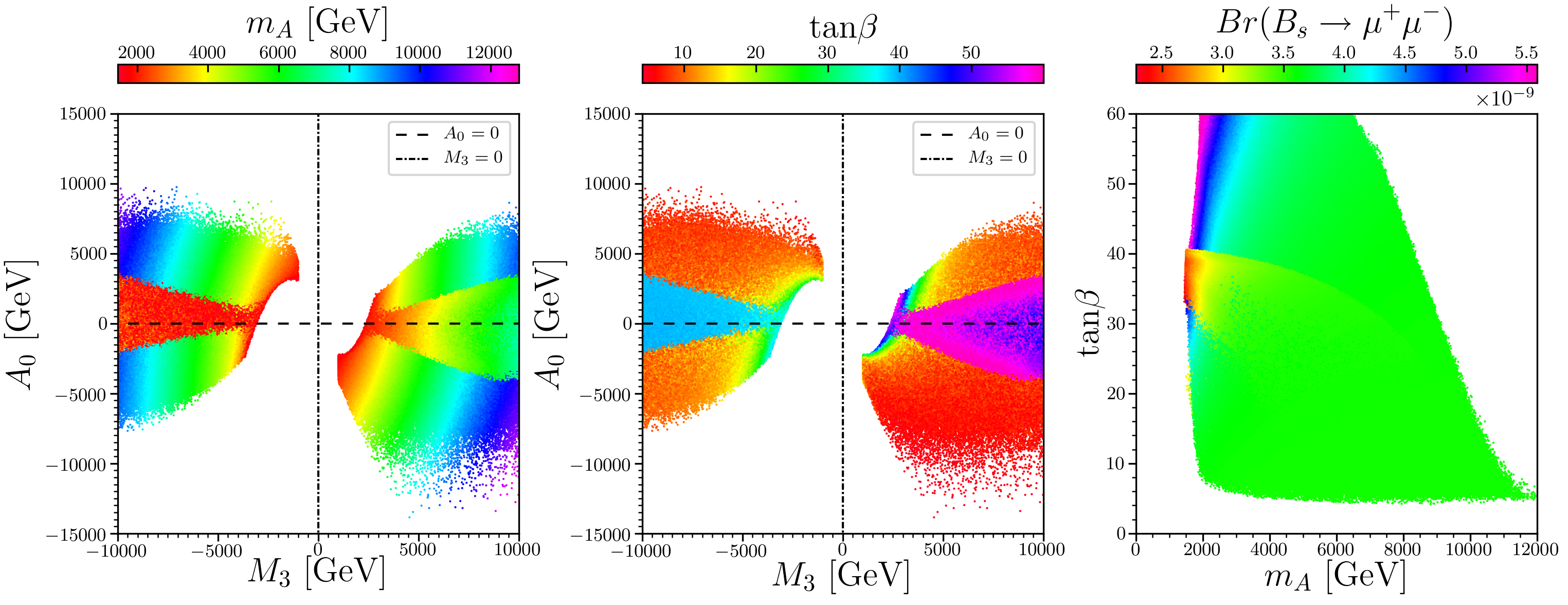}
\vspace*{-0.5cm}
\caption{Surviving samples in the $A_0$ versus $M_3$ (left and middle),
$\tan\beta$ versus the CP-odd Higgs mass $m_A$ (right) planes.
The colors indicate $m_A$ (left), $\tan\beta$ (middle), and $Br(B_s\to \mu^+\mu^-)$ (right), respectively.
\add{In these 2-dimension planes, larger-$\Delta a_\mu$ samples are projected on top of smaller ones.}
}
\label{fig2}
\end{figure}
In Fig.\ref{fig1}, we project the surviving samples on $A_0$ versus $M_3$ planes, with colors indicating the SM-like Higgs mass $m_h$ (left), the lighter stop mass $m_{\tilde{t}_1}$ (middle) and the higgsino mass parameter $\mu$ (right), respectively.
In Fig.\ref{fig2}, we project the surviving samples in the $A_0$ versus $M_3$ (left and middle), $\tan\beta$ versus the CP-odd Higgs mass $m_A$ (right) planes, with colors indicating $m_A$ (left), $\tan\beta$ (middle), and $Br(B_s\to \mu^+\mu^-)$ (right), respectively.
\add{In all these 2-dimension planes in Fig.\ref{fig1} and Fig.\ref{fig2}, larger-$\Delta a_\mu$ samples are shown on top of smaller ones.}
From the left plane of Fig.\ref{fig1}, left and middle planes of Fig.\ref{fig2}, we can see that \add{the larger-$\Delta a_\mu$ (or topside) samples on the $M_3-A_0$ planes can be sorted into four classes roughly:}
\begin{eqnarray}
  {\bf Class~A:} &&~~ 40\lesssim\tan\beta\lesssim60,
  ~|A_0/M_3|\lesssim 0.35, ~M_3>0, ~m_A\lesssim7\TeV; \nonumber\\
  {\bf Class~B:} &&~~ 35\lesssim\tan\beta\lesssim40,
  ~|A_0/M_3|\lesssim 0.35, ~M_3<0, ~m_A\lesssim3\TeV; \nonumber\\
  {\bf Class~C:} &&~~ \tan\beta\gtrsim15,
  ~|A_0/M_3|\gtrsim 0.35, ~m_A\lesssim4\TeV; \nonumber\\
  {\bf Class~D:} &&~~ \tan\beta\lesssim15,
  ~|A_0/M_3|\gtrsim 0.35, ~m_A\gtrsim4\TeV.
  \label{sort}
\end{eqnarray}
From the middle and right planes of Fig.\ref{fig1}, we can see that the lighter stop mass $m_{\tilde{t}_1}$ and parameter $\mu$ are mainly determined by $M_3$ and $A_0$.
Since $A_0,M_3\gg M_0,M_1,M_2$ and $\tan\beta \gg 1$, according to correlations between the parameters at SUSY scale and GUT scale are \add{listed} in Appendix A, we can have following approximations for most surviving samples:
\begin{eqnarray}
M_{\tilde{Q}_3} &\approx& M_{\tilde{U}_3} \approx 1.5 |M_3|, \nonumber\\
A_t &\approx& -1.0M_3-0.4A_0, \nonumber\\
\mu &=&
  \addii{
  \sqrt{\frac{M^2_{H_d}-M^2_{H_u}\tan^2\beta}{\tan^2\beta-1}-\frac{m_Z^2}{2}}\approx \left|M_{H_u}\right|
  }
  \approx \sqrt{0.91M_3^2 -0.18A_0 M_3 +0.09A_0^2},
\end{eqnarray}
Then we have
\begin{eqnarray}
m_{\tilde{t}_{1,2}} \approx \sqrt{(1.5M_3)^2\mp v|X_t|}
\end{eqnarray}
where $v=174\GeV$ is the Higgs VEV in the SM, and $X_t\equiv A_t-\mu/\tan\beta$.
The SM-like Higgs mass with one-loop correction of stops is given by
\begin{eqnarray}
 m_h^2 &=& m_Z^2\cos^2 2\beta +\frac{3}{4\pi^2}\frac{m_t^4}{v^2}
 \left[\log\frac{M_{\tilde{t}}^2}{m_t^2}
 +\frac{2X_t^2}{M_{\tilde{t}}^2}\left(1-\frac{X_t^2}{12M_{\tilde{t}}^2}\right)\right],
 \nonumber\\
 &\approx& m_Z^2 + \frac{3}{4\pi^2}\frac{m_t^4}{v^2}
 \times \left(2\log\frac{M_{\tilde{t}}}{m_t}+2x^2-\frac{1}{6}x^4\right)
\label{mhiggs}
\end{eqnarray}
where $x\equiv |X_t/M_{\tilde{t}}|$ and $M_{\tilde{t}}\equiv \sqrt{m_{\tilde{t}_1} m_{\tilde{t}_2}}$.
For the large-$\tan\beta$ samples \add{in Class A and B}, with $\tan\beta\gg1$, $|A_0/M_3|\lesssim0.35$, and farther approximating in Eq.(\ref{mhiggs}), we can get the Higgs mass can be mainly determined by $M_3$, \add{and when $|M_3|\thicksim 5\TeV$ the SM-like Higgs can get to 125 GeV with 1-loop stop corrections}.
For \add{moderate/large-$\tan\beta$ samples in Class C}, Higgs mass are mainly determined by both $M_3$ and $A_0$.
While for small-$\tan\beta$ samples \add{(Class D)}, $\tan\beta$ can also play a part in determining the SM-like Higgs mass.
\add{These are some of our new findings for larger-$\Delta a_\mu$ samples in this work.}

\addii{Just like the parameter point shown in Eqs.(\ref{P4MHu}, \ref{P4MHd}) in Appendix A,
for the samples in Class C and D ($|A_0/M_3|\gtrsim 0.35$ and $\tan\beta\gtrsim5$),
at $M_{\rm SUSY}$ scale we have $|M^2_{H_d}|\ll|M^2_{H_u}|$ and thus  the CP-odd Higgs mass
can be approximately given as}
\begin{eqnarray}
 m_A
 \addii{ =\sqrt{\frac{\tan^2\beta+1}{\tan^2\beta-1}\left(M^2_{H_d}-M^2_{H_u}\right)-m_Z^2}
 ~\approx \left|M_{H_u}\right|
 ~\approx \mu}.
\end{eqnarray}
\addii{While like the point shown in Eqs.(\ref{P7MHu}, \ref{P7MHd})  in Appendix A,
for the samples in Class A and B ($|A_0/M_3|\lesssim 0.35$ and $\tan\beta\gtrsim35$),
at $M_{\rm SUSY}$ scale $M^2_{H_d}$ can be comparable with $M^2_{H_u}$ and thus the
CP-odd Higgs mass can be much smaller than the parameter $\mu$, especially for
Class B where $M^2_{H_d}$ can be quite close to  $M^2_{H_u}$.}
\addii{We can see these characteristics jointly from the right plane of Fig.\ref{fig1},
the left and middle planes of Fig.\ref{fig2}.}
In SUSY models, we have the following equation for $B_s\to\mu^+\mu^-$ branch ratio
\begin{eqnarray}
 Br(B_s\to\mu^+\mu^-) \propto \frac{m_t^4 \mu^2 A_t^2 \tan^6\beta }{m_A^4 m_{\tilde{t}}^4}.
\end{eqnarray}
However, combing the three planes in Fig.\ref{fig2} we notice that the large-$\tan\beta$ \add{($\gtrsim 30$)} samples with \add{positive $M_3$ and} small-$m_A$ (2-3 TeV) predict large $B_s\to \mu^+\mu^-$ ratios, while the moderate-$\tan\beta$ \add{($35\thicksim40$)} samples with a negative $M_3$ \add{and a small $m_A$ (2-3 TeV)} predict small ratios.
\add{These positive- and negative-$M_3$ samples have different behaviors on $B_s\to \mu^+\mu^-$, which is another
new finding in our work. }

\begin{figure}[htb]
  \centering
\includegraphics[width=16cm]{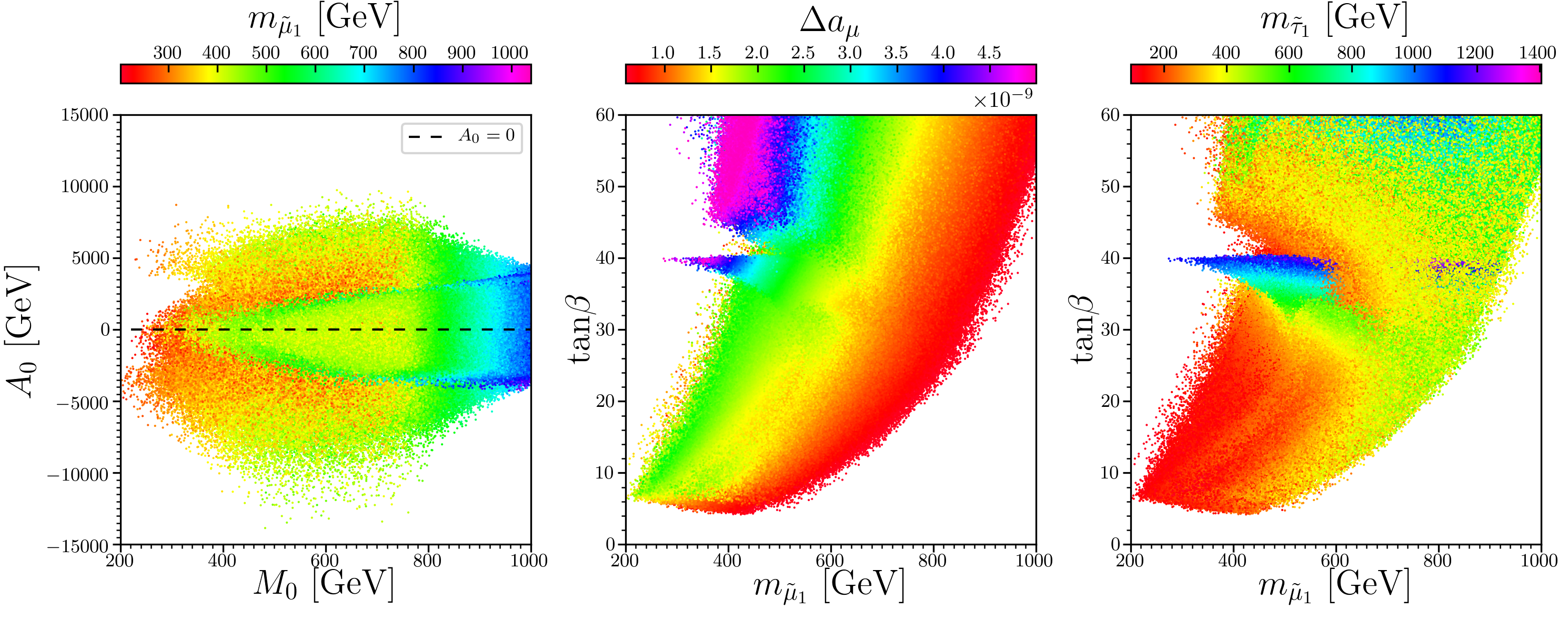}
\vspace*{-0.5cm}
\caption{Surviving samples in the $A_0$ \addii{versus} $M_0$ (left), and $\tan\beta$ \addii{versus} the lighter smuon
mass $m_{\tilde{\mu}_1}$ (middle and right) planes, with colors indicating
$m_{\tilde{\mu}_1}$ (left), SUSY contributions to muon g-2 $\Delta a_{\mu}$ (middle),
and the lighter stau mass $m_{\tilde{\tau}_1}$ (right), respectively.
\add{In these 2-dimension planes, larger-$\Delta a_\mu$ samples are projected on top of smaller ones.}
}
\label{fig3}
\end{figure}
In Fig.\ref{fig3}, we project surviving samples in the $A_0$ \addii{versus} $M_0$ (left), and $\tan\beta$ \addii{versus} the lighter smuon mass $m_{\tilde{\mu}_1}$ (middle and right) planes, with colors indicating $m_{\tilde{\mu}_1}$ (left), SUSY contributions to muon g-2 $\Delta a_{\mu}$ (middle), and the lighter stau mass $m_{\tilde{\tau}_1}$ (right), respectively.
\add{In these three 2-dimension planes, larger-$\Delta a_\mu$ samples are also shown on top of smaller ones.}
From the middle plane in Fig.\ref{fig3}, we can see that the muon g-2 anomaly can be solved in our scenario.
In fact, light smuon and large $\tan\beta$ can give a sizable contribution to $\Delta a_\mu$ with positive $\mu$ in MSSM.
Combined with Fig.\ref{fig2}, we can see that \add{the moderate-$\tan\beta$ ($35\thicksim40$) samples with negative-$M_3$} and predicting small $Br(B_s\to\mu^+\mu^-)$ can contribute sizably to $\Delta a_\mu$ \add{for light $\tilde{\mu}_1$ ($250\thicksim450$ GeV), but with heavy $\tilde{\tau}_1$ ($\gtrsim 1\TeV$) because of the exotic tuning among GUT parameters. This is also a new finding in this work.}
From the right plane we can know the light-$\tilde{\mu}_1$ and moderate-$\tan\beta$ \add{($10\thicksim45$) regions with positive $M_3$} are 
\add{missed} mainly because of the lower bounds of stau mass $m_{\tilde{\tau}_1}$.
\addii{The confusing missed part was shown in a figure in Ref.\cite{nugm-mug2}
but without an explanation, while we give a clear interpretation here in this work.}

\begin{figure}[htb]
  \centering
\includegraphics[width=16cm]{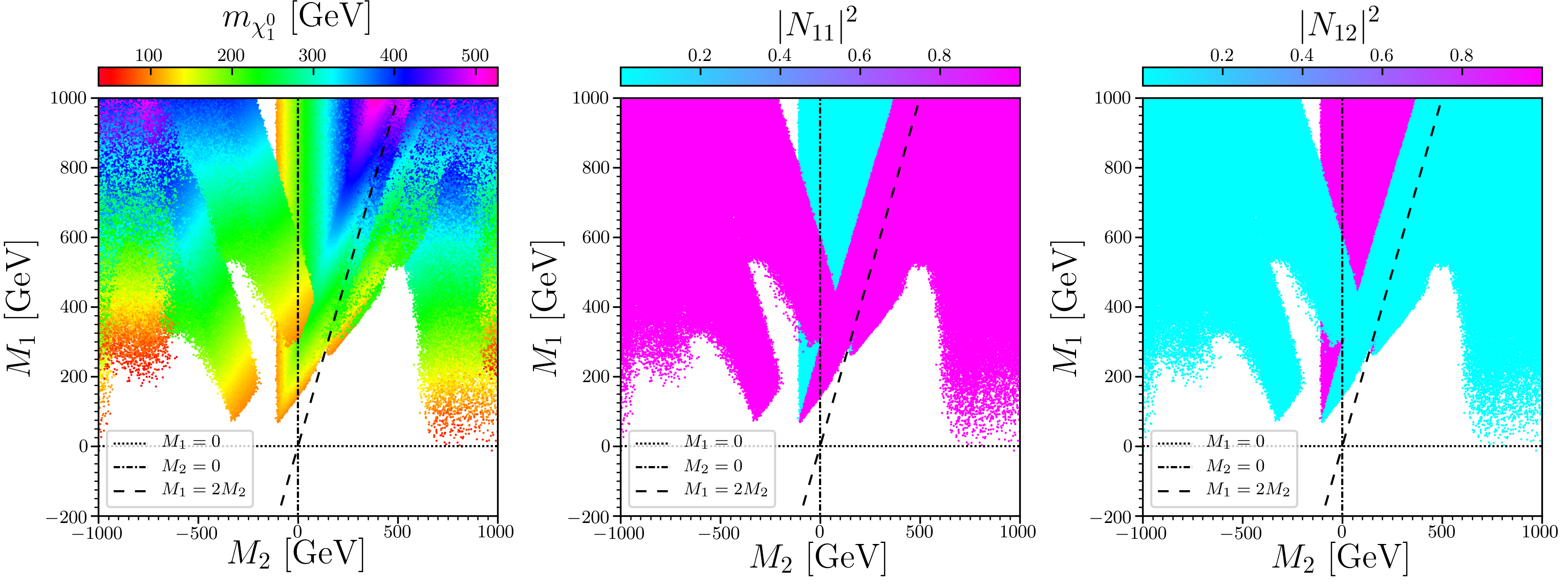}
\\ \includegraphics[width=16cm]{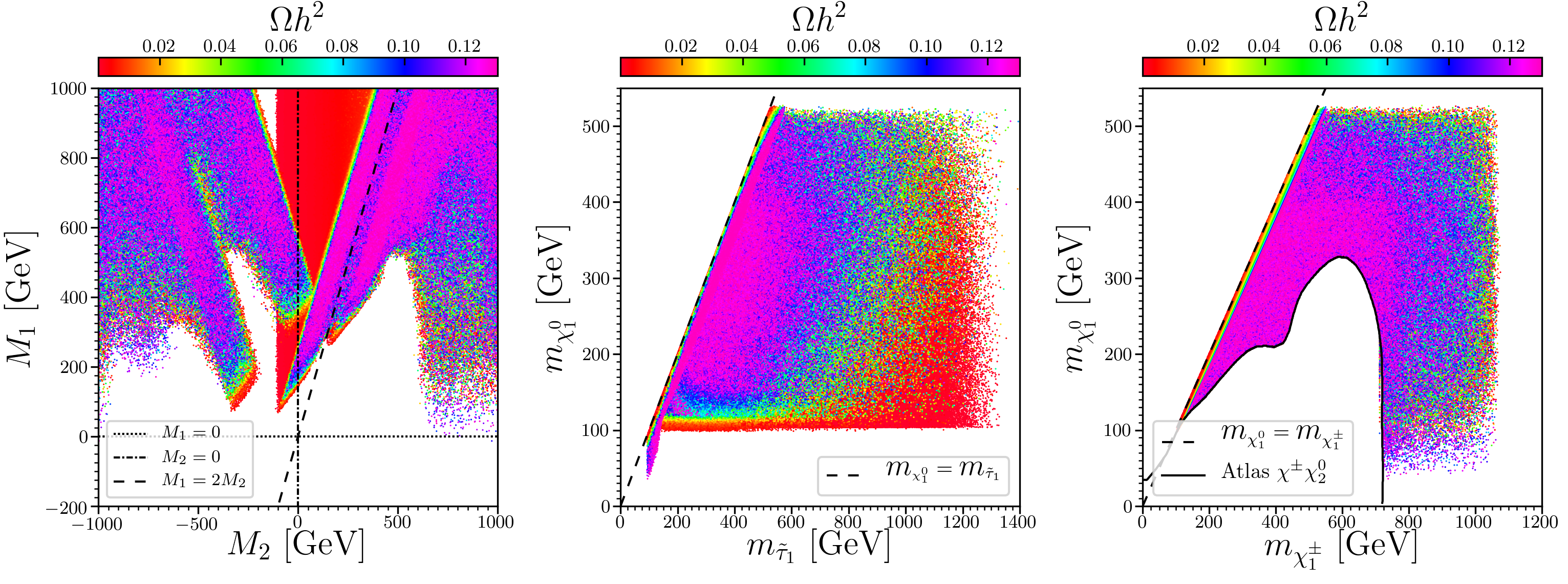}
\vspace*{-0.5cm}
\caption{
Surviving samples in the $M_1$ versus $M_2$ (upper 3 and lower left\add{, where $M_{1,2}$ are both defined at GUT scale}), LSP dark matter mass $m_{\chi^0_1}$ versus the lighter stau mass $m_{\tilde{\tau}_1}$ (lower middle), and $m_{\chi^0_1}$ versus the lighter chargino mass $m_{\chi^\pm_1}$ (lower right) planes.
Colors indicate $m_{\chi^0_1}$ (upper left), bino component in LSP (upper middle), wino component (upper right), and LSP relic density ($\Omega h^2$), respectively.
\add{In all these six 2-dimension planes, larger-$\Omega h^2$ samples are projected on top of smaller ones.}
\add{In the lower right planes, the black solid curve indicate the constraint of searching for $\chi^\pm\chi^0_2$ associate production in final states with tau leptons by Atlas collaboration at the 13 TeV LHC \cite{1708-a-ew}.}
}
\label{fig4}
\end{figure}
When $|M_3|\gg |M_1|, |M_2|$ \add{($M_{1,2,3}$ are defined at GUT scale, and $M_{1,2,3}^{\rm SUSY}$ hereafter are defined at $M_{\rm SUSY}$ scale)}, the RGE running of $M_3$ can have a visible influence on $M_1$ and $M_2$.
We checked in our samples that
\begin{eqnarray}
 {\rm when}~ M_3>0,
 &&~~-200 \lesssim M_2^{\rm SUSY}-0.85 M_2 \lesssim 0 \GeV,
 \nonumber \\
 &&~~-80 \lesssim M_1^{\rm SUSY}-0.45 M_1 \lesssim 0 \GeV;
 \nonumber \\
 {\rm when}~ M_3<0,
 &&~~~~~~~0 \lesssim M_2^{\rm SUSY}-0.85 M_2 \lesssim 200 \GeV,
 \nonumber \\
 &&~~~~~~~0 \lesssim M_1^{\rm SUSY}-0.45 M_1 \lesssim 80 \GeV,
\end{eqnarray}
\add{which can also be interpreted with the equations in Appendix A.}
Since \add{$M_1^{\rm SUSY}$ and $M_2^{\rm SUSY}$} are both in the diagonal position of the neutralino mass matrix, and \add{$\mu\gg |M_1^{\rm SUSY}|, |M_2^{\rm SUSY}|$}, the lightest neutralino (LSP) are either bino or wino, with nearly no mixing between them.
The above discussions can be shown on the top planes in Fig.\ref{fig4}.
\add{In all these 2-dimension planes in Fig.\ref{fig4}, larger relic density samples are also shown on top of smaller ones.}
From the bottom left plane, we can see that only bino-like LSP can generate enough relic density.
While the wino neutralino always coannihilates with the wino charginos, the relic density is always too small to account for full abundance.
We checked that for bino LSP,  all the decay modes of wino neutralino $\chi^0_2$ and chargino $\chi^\pm_1$ contain a $\tau$ or $\tilde{\tau}_1$ final state. Thus, from the bottom right plane, we can see the searches for EW gauginos at the LHC set important constraints to the model.
\add{From the approximate equations and Fig.\ref{fig4}, we found a correlation between $M_{1,2}^{\rm SUSY}$ and $M_3$,
and its influence on dark matter composition, especially the deviation of boundaries from $|M_1/M_2|=2$ at the GUT scale.}

\begin{table}[b]
\vspace*{-0.5cm}
\centering
\caption{The main annihilation channels and their relative contributions to $\langle\sigma v\rangle$ for the 7 benchmark points.}
\label{table2}
\begin{spacing}{1.05}
\begin{tabular}{|c|c|c|c|}
\hline\hline
\makecell[tl]{
\qquad P1, $\tilde{\tau}_1$ coann.\\ \hline
$ \chi^0_1 \tilde{\tau}_1 \to \tau h$, $93.6\%$ \\
$ \chi^0_1 \chi^0_1 \to \tau \tau$, $3.2\%$ \\
$ \chi^0_1 \tilde{\tau}_1 \to W^- \nu_\tau$, $1.6\%$ \\
$ \chi^0_1 \tilde{\tau}_1 \to Z \tau$, $1.2\%$ \\
\\ \hline\hline
\qquad P4, $\tilde{\tau}_1$ exch. \\ \hline
$ \chi^0_1 \chi^0_1 \to \tau \tau$, $99.1\%$ \\
\\ \hline\hline
\quad ~P5, $\tilde{\tau}_1$ hybrid3 \\ \hline
$ \chi^0_1 \tilde{\tau}_1 \to \tau h$, $89.9\%$ \\
$ \chi^0_1 \chi^0_1 \to \tau \tau$, $8.1\%$ \\
\\ \hline\hline
\quad ~P6, $\tilde{\tau}_1$ hybrid3 \\ \hline
$ \chi^0_1 \tilde{\tau}_1 \to \tau h$, $89.7\%$ \\
$ \chi^0_1 \chi^0_1 \to \tau \tau$, $3.3\%$ \\
$ \chi^0_1 \tilde{\tau}_1 \to W^- \nu_\tau$, $3.1\%$ \\
$ \chi^0_1 \tilde{\tau}_1 \to Z \tau$, $2.8\%$
}
&
\makecell[tl]{
\qquad P2, $\chi^+_1$ coann.\\ \hline
$ \chi^+_1 \chi^0_2 \to u \bar{d}$, $11.9\% $ \\
$ \chi^+_1 \chi^0_2 \to c \bar{s}$, $11.9\% $ \\
$ \chi^+_1 \chi^0_2 \to t \bar{b}$, $11.4\% $ \\
$ \chi^+_1 \chi^+_1 \to W^+ W^+$, $8.7\% $ \\
$ \chi^0_2 \chi^0_2 \to W^+ W^-$, $8.7\% $ \\
$ \chi^+_1 \chi^0_2 \to Z W^+$, $8.0\% $ \\
$ \chi^+_1 \chi^-_1 \to Z Z$, $5.2\% $ \\
$ \chi^+_1 \chi^-_1 \to W^+ W^-$, $5.0\% $ \\
$ \chi^+_1 \chi^-_1 \to \gamma Z$, $3.0\% $ \\
$ \chi^+_1 \chi^-_1 \to s \bar{s}$, $3.0\% $ \\
$ \chi^+_1 \chi^-_1 \to d \bar{d}$, $3.0\% $ \\
$ \chi^+_1 \chi^-_1 \to b \bar{b}$, $3.0\% $ \\
$ \chi^+_1 \chi^-_1 \to u \bar{u}$, $3.0\% $ \\
$ \chi^+_1 \chi^-_1 \to c \bar{c}$, $3.0\% $ \\
$ \chi^+_1 \chi^-_1 \to t \bar{t}$, $2.7\% $ \\
$ \chi^+_1 \chi^0_2 \to \gamma W^+$, $2.0\% $
}
& \makecell[tl]{
\qquad P3, hybrid2\\ \hline
$ \tilde{\tau}_1 \tilde{\tau}_1 \to h h$, $18.4\% $ \\
$ \chi^0_1 \tilde{\tau}_1 \to \tau h$, $9.7\% $ \\
$ \chi^+_1 \chi^0_2 \to u \bar{d}$, $5.7\% $ \\
$ \chi^+_1 \chi^0_2 \to c \bar{s}$, $5.7\% $ \\
$ \chi^+_1 \chi^0_2 \to t \bar{b}$, $5.4\% $ \\
$ \chi^0_2 \chi^0_2 \to W^+ W^-$, $4.1\% $ \\
$ \chi^+_1 \chi^+_1 \to W^+ W^+$, $4.1\% $ \\
$ \chi^+_1 \chi^0_2 \to Z W^+$, $3.8\% $ \\
$ \tilde{\tau}_1 \tilde{\tau}_1 \to Z Z$, $2.5\% $ \\
$ \chi^+_1 \chi^-_1 \to Z Z$, $2.5\% $ \\
$ \chi^+_1 \chi^-_1 \to W^+ W^-$, $2.4\% $ \\
$ \chi^+_1 \tilde{\tau}_1 \to \nu_\tau h$, $2.1\% $ \\
$ \chi^+_1 \chi^0_1 \to \nu_\tau \tau^+$, $2.1\% $ \\
$\chi^+_1 \chi^0_1 \to \nu_e e^+$, $1.9\% $ \\
$\chi^+_1 \chi^0_1 \to \nu_\mu \mu^+$, $1.9\% $ \\
$\tilde{\tau}_1 \tilde{\tau}_1 \to W^+ W^-$, $1.8\% $ \\
$\chi^0_1 \tilde{\tau}_1 \to Z \tau$, $1.7\% $ 
}
& \makecell[tl]{
\qquad P7, $\tilde{\ell}$ ann.\\ \hline
$ \tilde{\mu} \tilde{\nu}_e \to e \nu_\mu$, $11.3\% $ \\
$ \tilde{e} \tilde{\nu}_\mu \to \nu_e \mu$, $11.3\% $ \\
$ \tilde{\mu} \tilde{\nu}_e \to \nu_e \mu$, $4.7\% $ \\
$ \tilde{e} \tilde{\nu}_\mu \to e \nu_\mu$, $4.7\% $ \\
$ \tilde{\nu}_e \tilde{\bar{\nu}}_e \to W^+ W^-$, $4.6\% $ \\
$ \tilde{\nu}_\mu \tilde{\bar{\nu}}_\mu \to W^+ W^-$, $4.6\% $ \\
$ \tilde{\nu}_e \tilde{\bar{\nu}}_e \to Z Z$, $4.0\% $ \\
$ \tilde{\nu}_\mu \tilde{\bar{\nu}}_\mu \to Z Z$, $4.0\% $ \\
$ \tilde{e} \tilde{\bar{\nu}}_e \to \gamma W^-$, $2.4\% $ \\
$ \tilde{\mu} \tilde{\bar{\nu}}_\mu \to \gamma W^-$, $2.4\% $ \\
$ \tilde{\nu}_e \tilde{\nu}_\mu \to \nu_e \nu_\mu$, $2.3\% $ \\
$ \tilde{\nu}_e \tilde{\nu}_e \to \nu_e \nu_e$, $2.3\% $ \\
$ \tilde{\nu}_\mu \tilde{\nu}_\mu \to \nu_\mu \nu_\mu$, $2.3\% $ \\
$ \tilde{e} \tilde{\bar{\nu}}_e \to Z W^-$, $2.0\% $ \\
$ \tilde{\mu} \tilde{\bar{\nu}}_\mu \to Z W^-$, $2.0\% $ \\
$ \tilde{e} \tilde{e} \to W^+ W^-$, $2.0\% $ \\
$ \tilde{\mu} \tilde{\mu} \to W^+ W^-$, $2.0\% $ 
}
\\ \hline \hline
\end{tabular}
\end{spacing}
\end{table}
From the bottom middle and right planes, we can also glimpse the annihilation mechanisms of bino-like LSP in our model.
We checked that for samples predicting the right relic density \add{($0.107<\Omega h^2<0.131$)}, there are mainly five single mechanisms \footnote{\add{Notice that unlike that in CMSSM, we do not have surviving samples with stop as the next-to-lightest SUSY particle (NLSP), and stop coannihilation mechanism in our NUGM extension, which is because in our scenario we can have lighter bino-like neutralino and wino-like charginos for NUGM, and lighter sleptons to solve the muon g-2.}}
and several combined ones :
stau exchange ($\chi^0_1 \chi^0_1 \to \tau^+\tau^-$),
stau coannihilation ($\chi^0_1 \tilde{\tau}_1^\pm \to \tau^\pm h/Z, \, W^\pm\nu_\tau$
),
chargino coannihilation ($\chi^0_1 \chi^\pm_1 \to \ell^\pm \nu_\ell$)
,
stau annihilation ($\tilde{\tau}^+_1 \tilde{\tau}^-_1 \to hh$
),
other slepton annihilation ($\tilde{\ell}/\tilde{\nu}_\ell \tilde{\ell}/\tilde{\nu}_\ell \to XY$)
.
Thus we sort our surviving samples into six classes by judging if it is a single or a combined mechanism:
\vspace*{-0.6cm}
\begin{eqnarray}
  {\rm \tilde{\tau}_1~ exchange:}
  &&\quad m_{\tilde{\tau}_1} < 200 \GeV, \nonumber\\
  {\rm \tilde{\tau}_1~ coannihilation:}
  &&\quad \frac{m_{\tilde{\tau}_1}}{m_{\chi^0_1}}<1.2,
  ~~\frac{m_{\chi^{\pm}_1}}{m_{\chi^0_1}}>1.2, \nonumber\\
  \chi^{\pm}_1~ {\rm coannihilation:}
  &&\quad \frac{m_{\tilde{\tau}_1}}{m_{\chi^0_1}}>1.2,
  ~~\frac{m_{\chi^{\pm}_1}}{m_{\chi^0_1}}<1.2, \nonumber\\
  {\rm hybrid2:}
  &&\quad \frac{m_{\tilde{\tau}_1}}{m_{\chi^0_1}}<1.2,
  ~~\frac{m_{\chi^{\pm}_1}}{m_{\chi^0_1}}<1.2, \nonumber\\
  {\rm \tilde{\tau}_1~ hybrid3:}
  &&\quad \frac{m_{\tilde{\tau}_1}}{m_{\chi^0_1}}>1.2,
  ~~\frac{m_{\chi^{\pm}_1}}{m_{\chi^0_1}}>1.2,
  ~~200 < m_{\tilde{\tau}_1} < 400 \GeV, \nonumber\\
  {\rm \tilde{\ell}~ annihilation:}
  &&\quad \frac{m_{\tilde{\tau}_1}}{m_{\chi^0_1}}>1.2,
  ~~\frac{m_{\chi^{\pm}_1}}{m_{\chi^0_1}}>1.2,
  ~~m_{\tilde{\tau}_1} > 400 \GeV.
\end{eqnarray}
For the hybrid2 samples, the dominated mechanism is a combined one by $\tilde{\tau}_1$ coannihilation and $\chi^\pm_1$ coannihilation;
while for stau hybrid3, it is combined by $\tilde{\tau}_1$ exchange, $\tilde{\tau}_1$ coannihilation and $\tilde{\tau}_1$ annihilation, and the heavier $\tilde{\tau}_1$, the more annihilation and the less exchange;
but when $\tilde{\tau}_1$ are heavier than 400 GeV, the dominated mechanism becomes other sleptons coannihilation, which is very complex in income and outcome particles.
In Tab.\ref{table2}, we give the detail annihilation information for 7 benchmark points.
For each point, we list its main annihilation channels and the relative contributions ($>1.5\%$) to $\langle\sigma v\rangle$.
For completeness, we list the other information for the benchmark points in Tab.\ref{table3} in Appendix B.
\add{In this work, we show in detail the various annihilation mechanisms of DM, which is not done
in Ref.\cite{nugm-mug2}.
And our findings is some different from these in Ref.\cite{pMSSM10} for NUGM version of pMSSM.
We do not have annihilation mechanisms of A/H funnel, focus point, and stop coannihilation, since our $H/A,\mu,\tilde{t}_1$ are much heavier, but we have stau exchange which may be omitted in Ref.\cite{pMSSM10}.}

\begin{figure}[!htb]
  \centering
\includegraphics[width=10.6cm]{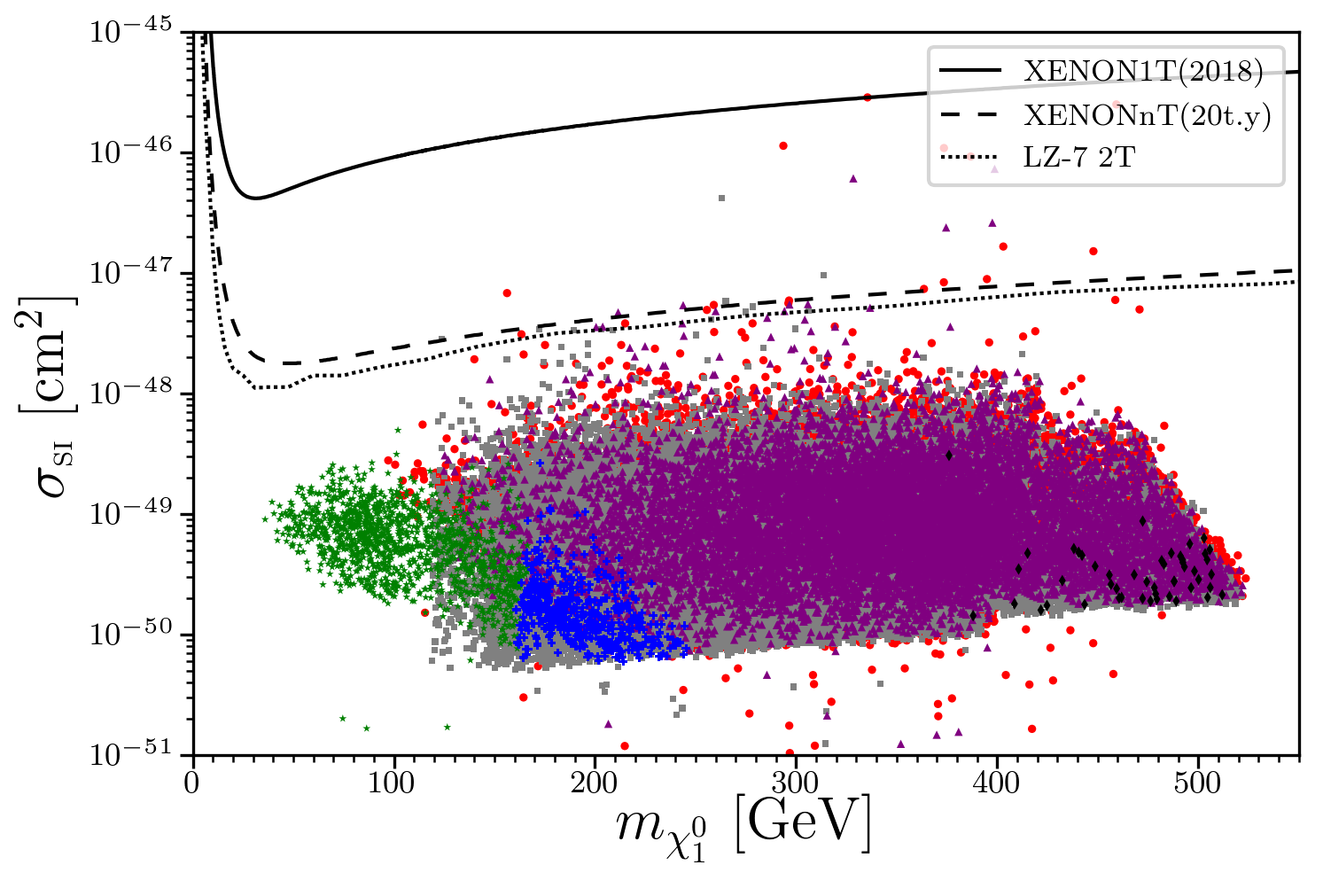}
\vspace*{-0.5cm}
\caption{Surviving samples with sufficient dark matter relic density \add{($0.107<\Omega h^2<0.131$)} in the spin-independent DM-nucleon
cross section $\sigma_{\rm SI}$ \add{(original values without being rescaled by $\Omega/\Omega_0$)} \addii{versus} LSP DM mass $m_{\chi^0_1}$.
The limits of XENON1T in 2018, the future detection sensitivity of XENONnT and LUX-ZEPLIN (LZ-7 2T)
are shown by real, dashed and dotted curves respectively.
Different annihilation scenarios of samples are also shown by different
symbols with different colors:
$\tilde{\tau}_1$ exchange by green star `$\green \star$', $\tilde{\tau}_1$ hybrid3 by blue cross `$\blue +$',
$\tilde{\tau}_1$ coannihilation by red bullet `$\red\bullet$',
$\chi^{\pm}$ coannihilation by purple triangle `$\purple \blacktriangle$',
hybrid2 by gray square `$\gray \blacksquare$',
and $\tilde{\ell}$ annihilation by black lozenge `$\blacklozenge$'.
}
\label{fig5}
\end{figure}
In Fig.\ref{fig5}, we show the six classes of samples with sufficient relic density \add{($0.107<\Omega h^2<0.131$)} on the plane of
SI DM-nucleon cross section $\sigma_{\rm SI}$ \add{(original values without being rescaled by $\Omega/\Omega_0$)} versus LSP mass $m_{\chi^0_1}$.
We can see that, most of the samples predict small $\sigma_{\rm SI}$, which are over one order of magnitude lower than the future detection accuracy of LZ and XENONnT experiments.
\add{The $\sigma_{\rm SI}$ are smaller than these in Ref.\cite{nugm-mug2}, because we required sufficient relic density for these samples, and we checked that $\sigma_{\rm SI}$ can be larger for samples with insufficient relic density.}
However, a few samples corresponding to $\tilde{\tau}_1$ coannihilation, $\chi^\pm_1$ coannihilation and hybrid2 can be covered by the two detectors, with the LSP mass at about 200-400 GeV.
\add{It is because these samples have large percentages of coannihilation channels contributing to the DM relic density, which is also a new finding in this work. }

\section{Summary and Conclusions}
We propose to generate non-universal gaugino masses in SU(5) GUT with the generalized Planck-scale mediation
SUSY breaking mechanism, in which the non-universality arises from proper wavefunction normalization with lowest component
VEVs of various high dimensional representations of the Higgs fields of SU(5) and an unique F-term VEV by the singlet.
 Different predictions on gaugino mass ratios with respect to widely studied scenarios are given.
The gluino-SUGRA-like scenarios, where gluinos are much heavier than winos, bino and universal scalar masses,
can be easily realized with appropriate combinations of such high-representation Higgs fields.
With six GUT-scale free parameters in our scenario, we can solve 
\add{elegantly}
\add{the tension in mSUGRA between the muon g-2 and other constraints including the dark matter relic density and the direct sparticle search bounds from the LHC.}

Taking into account the current constraints, we performed a scan and obtained the following observations:
\bit
\vspace{-0.2cm}\setlength{\itemsep}{-2pt}
\item \add{The large-$\tan\beta$ ($\gtrsim35$) samples with a moderate $M_3$ ($\thicksim 5\TeV$), a small $|A_0/M_3|$ ($\lesssim0.35$) and a small $m_A$ ($\lesssim 4\TeV$) are favoured to generate a 125 GeV SM-like Higgs and predict a large muon g-2, while the stops mass and $\mu$ parameter, which are mainly determined by $|M_3|$ ($\gg M_0,|M_1|,|M_2|$), can be about 6 TeV.}
\item \add{The moderate-$\tan\beta$ ($35\thicksim40$) samples with a negative $M_3$ can have a light smuon ($250\thicksim450$ GeV) but a heavy stau ($\gtrsim 1\TeV$), which predict a large muon g-2 but small $Br(B_s\to\mu^+\mu^-)$.}
\item The lightest neutralino can be as light as 100 GeV, which can predict a right relic abundance if it is bino-like and a much smaller relic density if it is wino-like.
\item To obtain the right DM relic density, the annihilation mechanisms should be stau exchange, stau coannihilation, chargino coannihilation, slepton annihilation and the combination of two or three of them;
\item To obtain the right DM relic density, the spin-independent DM-nucleon cross section is typically much smaller than the present bounds of XENON1T 2018, and an order of magnitude lower than the future detection sensitivity of LZ and XENONnT experiments.
\eit

\section*{Acknowledgement}
This work was supported in part by the National Natural Science Foundation of China (NNSFC)
under grant Nos. 11605123, 11675147, 11547103, 11547310, 11675242, 11851303, 11821505, 
by the Innovation Talent Project of Henan Province under grant number 15HASTIT017,
by the Young Core Instructor Foundation of Henan Education Department,
by Peng-Huan-Wu Theoretical Physics Innovation Center (11747601),
by the CAS Center for Excellence in Particle Physics (CCEPP),
by the CAS Key Research Program of Frontier Sciences
and by a Key R\&D Program of Ministry of Science and Technology under number 2017YFA0402200-04.
JZ also thanks Prof. Stephen P. Martin for helpful discussion,
thanks the support of the China Scholarship Council (CSC) under Grant No.201706275160
while he was working at the University of Chicago as a visiting scholar,
and thanks the support of the US National Science Foundation (NSF) under Grant No. PHY-0855561 while he was working at Michigan State University.

\section*{Appendix A: Correlations between the parameters at SUSY scale and GUT scale }
We show the correlations between the parameters at soft SUSY scale and GUT scale.
For the benchmark point P4, the GUT scale is calculated to be $M_{\rm GUT}=1.27\times 10^{16} \GeV$.
Then we use two-loop RGEs to run the parameters from GUT scale to the SUSY scale, which we choose as $M_{\rm SUSY}=4\TeV$.
We repeat this process over 15 times by slightly changing the following one or two parameters excluding $\tan\beta$ at GUT scale each time
\begin{eqnarray}
p_{j,k}^{\rm GUT}=A_0,M_0,M_1,M_2,M_3.
\end{eqnarray}
For the linear-correlation parameters
\begin{eqnarray}
p_i^{\rm SUSY}=A_t,A_\tau,A_\mu,M^{\rm SUSY}_1,M^{\rm SUSY}_2,M^{\rm SUSY}_3
\end{eqnarray}
we calculate the coefficients by
\begin{eqnarray}
C_{ij}=\frac{\Delta p_i^{\rm SUSY}}{\Delta p_j^{\rm GUT}}.
\end{eqnarray}
For the quadratic-correlation parameters
\begin{eqnarray}
p_i^{\rm SUSY}=\mu^2, M^2_{H_u}, M^2_{H_d}, M^2_{Q_3}, M^2_{U_3}, M^2_{L_3}, M^2_{E_3}, M^2_{L_2}, M^2_{E_2}
\end{eqnarray}
we calculate the coefficients by
\begin{eqnarray}
C_{ijk(k\geqslant j)}=\frac{n\Delta p_i^{\rm SUSY}}{\Delta p_j^{\rm GUT} \Delta p_k^{\rm GUT}}
\quad (n=2 {\rm ~for~} k=j, ~n=1 {\rm ~for~} k>j),
\end{eqnarray}
\addii{which can be written in a 5$\times$5 triangular-matrix for each parameter.
In the following equations, we list the coefficients for the benchmark point P4
in Class C and D in Eq.(\ref{sort}), and P7 in Class A and B in Eq.(\ref{sort}).
We checked that these coefficients coincide approximately with our parameter-running results in \textsf{NMSSMTools-5.2.0}.
Most of these equations (except $M^2_{H_d}$ and $m^2_A$, for example) can be generalized roughly to other surviving samples in their represented classes, because all of them satisfy $\tan\beta\gg 1$.}
However, most coefficients will change a lot if one change the SUSY scale too much, e.g., to $M_{\rm SUSY}=400 \GeV$ as in Ref.\cite{nugm-Martin}.
These equations are given as follows:

\begin{spacing}{1.0}
For Benchmark Point P4 (with $\tan\beta=20.8$ fixed and on behalf of Class C and D),
\begin{eqnarray}
 A_t &=& 0.40A_0 -1.16M_3 -0.04M_0 -0.04M_1 -0.18M_2 \label{P4At}\\ 
 A_\tau &=& 0.93A_0 -0.40M_2 -0.14M_1 +0.028M_3 \label{P4Atau}\\ 
 A_\mu &=& 0.96A_0 -0.41M_2 -0.14M_1 +0.028M_3 \label{P4Amu}\\ 
 M^{\rm SUSY}_1 &=&  0.46M_1  -0.007M_3 \label{P4M1}\\ 
 M^{\rm SUSY}_2 &=&  0.83M_2  -0.018M_3 \label{P4M2}\\ 
 M^{\rm SUSY}_3 &=&  1.92M_3  -0.08M_2 +0.04M_0 \label{P4M3}
\end{eqnarray}
\begin{eqnarray}
M_{\tilde{Q}_3}^2 &=&
  \left( \begin{array}{ccccc}
  A_0 & M_0 & M_1 & M_2 & M_3 \\
  \end{array} \right)
  \left( \begin{array}{ccccc}
   -0.067 & -0.29 & -0.01 &  0.04 &  0.09 \\
    \zro & -1.01 & -0.30 &  0.43 &  0.26 \\
    \zro &  \zro &  0.11 &  0.01 &  0.01 \\
    \zro &  \zro &  \zro &  1.05 & -0.51 \\
    \zro &  \zro &  \zro &  \zro &  2.15 \\
 \end{array} \right)
  \left( \begin{array}{c}
  A_0 \\ M_0 \\ M_1 \\ M_2 \\ M_3 \\
  \end{array} \right)
 \label{P4MQ3} 
\end{eqnarray}
\begin{eqnarray}
M_{\tilde{U}_3}^2 &=&
  \left( \begin{array}{ccccc}
  A_0 & M_0 & M_1 & M_2 & M_3 \\
  \end{array} \right)
  \left( \begin{array}{ccccc}
   -0.096 & -0.24 & -0.00 &  0.05 &  0.15 \\
    \zro & -0.99 & -0.25 &  0.35 &  0.22 \\
    \zro &  \zro &  0.13 &  0.01 &  0.00 \\
    \zro &  \zro &  \zro &  0.51 & -0.44 \\
    \zro &  \zro &  \zro &  \zro &  1.92 \\
 \end{array} \right)
  \left( \begin{array}{c}
  A_0 \\ M_0 \\ M_1 \\ M_2 \\ M_3 \\
  \end{array} \right)
\label{P4MU3}
\end{eqnarray}
\begin{eqnarray}
M_{H_u}^2 &=&
  \left( \begin{array}{ccccc}
  A_0 & M_0 & M_1 & M_2 & M_3 \\
  \end{array} \right)
\left( \begin{array}{ccccc}
   -0.102 &  0.20 &  0.02 &  0.03 &  0.22 \\
    \zro &  1.18 &  0.19 & -0.28 & -0.16 \\
    \zro &  \zro & -0.06 & -0.01 & -0.03 \\
    \zro &  \zro &  \zro & -0.23 &  0.18 \\
    \zro &  \zro &  \zro &  \zro & -0.91 \\
 \end{array} \right)
  \left( \begin{array}{c}
  A_0 \\ M_0 \\ M_1 \\ M_2 \\ M_3 \\
  \end{array} \right)
 \label{P4MHu}
\end{eqnarray}
\begin{eqnarray}
M_{H_d}^2 &=&
  \left( \begin{array}{ccccc}
  A_0 & M_0 & M_1 & M_2 & M_3 \\
  \end{array} \right)
\left( \begin{array}{ccccc}
   -0.027 &  0.03 &  0.00 &  0.01 &  0.04 \\
    \zro &  0.99 &  0.02 & -0.03 & -0.02 \\
    \zro &  \zro &  0.02 & -0.00 & -0.00 \\
    \zro &  \zro &  \zro &  0.31 &  0.00 \\
    \zro &  \zro &  \zro &  \zro & -0.09 \\
 \end{array} \right)
  \left( \begin{array}{c}
  A_0 \\ M_0 \\ M_1 \\ M_2 \\ M_3 \\
  \end{array} \right)
 \label{P4MHd}
\end{eqnarray}
\begin{eqnarray}
\mu^2 &=&
  \frac{M^2_{H_d}-M^2_{H_u}\tan^2\beta}{\tan^2\beta-1}-\frac{m_Z^2}{2}
  \nonumber\\
  &\approx&
  \left( \begin{array}{ccccc}
  A_0 & M_0 & M_1 & M_2 & M_3 \\
  \end{array} \right)
\left( \begin{array}{ccccc}
    0.102 & -0.20 & -0.02 & -0.03 & -0.22 \\
    \zro & -1.18 & -0.19 &  0.28 &  0.16 \\
    \zro &  \zro &  0.06 &  0.01 &  0.03 \\
    \zro &  \zro &  \zro &  0.24 & -0.18 \\
    \zro &  \zro &  \zro &  \zro &  0.91 \\
 \end{array} \right)
  \left( \begin{array}{c}
  A_0 \\ M_0 \\ M_1 \\ M_2 \\ M_3 \\
  \end{array} \right)
 \label{P4mu} 
\end{eqnarray}
\begin{eqnarray}
m_{A}^2 &=&
  \frac{\tan^2\beta+1}{\tan^2\beta-1}\left(M^2_{H_d}-M^2_{H_u}\right)-m_Z^2
  \nonumber\\
  &\approx&
  \left( \begin{array}{ccccc}
  A_0 & M_0 & M_1 & M_2 & M_3 \\
  \end{array} \right)
  \left( \begin{array}{ccccc}
    0.075 & -0.18 & -0.02 & -0.02 & -0.19 \\
    \zro & -0.19 & -0.17 &  0.25 &  0.15 \\
    \zro &  \zro &  0.09 &  0.01 &  0.02 \\
    \zro &  \zro &  \zro &  0.54 & -0.18 \\
    \zro &  \zro &  \zro &  \zro &  0.82 \\
 \end{array} \right)
  \left( \begin{array}{c}
  A_0 \\ M_0 \\ M_1 \\ M_2 \\ M_3 \\
  \end{array} \right)
 \label{P4MA}
\end{eqnarray}
\begin{eqnarray}
M_{\tilde{L}_3}^2 &=&
  \left( \begin{array}{ccccc}
  A_0 & M_0 & M_1 & M_2 & M_3 \\
  \end{array} \right)
  \left( \begin{array}{ccccc}
   -0.011 &  0.00 &  0.00 &  0.00 & -0.00 \\
    \zro &  0.96 &  0.00 &  0.00 & -0.00 \\
    \zro &  \zro &  0.03 & -0.00 & -0.00 \\
    \zro &  \zro &  \zro &  0.37 & -0.00 \\
    \zro &  \zro &  \zro &  \zro & -0.004 \\
 \end{array} \right)
  \left( \begin{array}{c}
  A_0 \\ M_0 \\ M_1 \\ M_2 \\ M_3 \\
  \end{array} \right)
 \label{P4ML3}
\end{eqnarray}
\begin{eqnarray}
M_{\tilde{L}_2}^2 &=&
  \left( \begin{array}{ccccc}
  A_0 & M_0 & M_1 & M_2 & M_3 \\
  \end{array} \right)
  \left( \begin{array}{ccccc}
    0.000 &  0.00 & -0.00 & -0.00 &  0.00 \\
    \zro &  0.99 &  0.00 &  0.00 & -0.00 \\
    \zro &  \zro &  0.03 & -0.00 & -0.00 \\
    \zro &  \zro &  \zro &  0.38 & -0.00 \\
    \zro &  \zro &  \zro &  \zro & -0.004 \\
 \end{array} \right)
  \left( \begin{array}{c}
  A_0 \\ M_0 \\ M_1 \\ M_2 \\ M_3 \\
  \end{array} \right)
 \label{P4ML2}
\end{eqnarray}
\begin{eqnarray}
M_{\tilde{E}_3}^2 &=&
  \left( \begin{array}{ccccc}
  A_0 & M_0 & M_1 & M_2 & M_3 \\
  \end{array} \right)
  \left( \begin{array}{ccccc}
   -0.022 &  0.00 &  0.00 &  0.01 & -0.00 \\
    \zro &  0.93 &  0.00 & -0.00 &  0.00 \\
    \zro &  \zro &  0.13 & -0.00 & -0.00 \\
    \zro &  \zro &  \zro & -0.01 &  0.00 \\
    \zro &  \zro &  \zro &  \zro & -0.000 \\
 \end{array} \right)
  \left( \begin{array}{c}
  A_0 \\ M_0 \\ M_1 \\ M_2 \\ M_3 \\
  \end{array} \right)
 \label{P4ME3}
\end{eqnarray}
\begin{eqnarray}
M_{\tilde{E}_2}^2 &=&
  \left( \begin{array}{ccccc}
  A_0 & M_0 & M_1 & M_2 & M_3 \\
  \end{array} \right)
  \left( \begin{array}{ccccc}
    0.000 &  0.00 & -0.00 &  0.00 & -0.00 \\
    \zro &  1.00 &  0.00 & -0.00 & -0.00 \\
    \zro &  \zro &  0.13 & -0.00 & -0.00 \\
    \zro &  \zro &  \zro & -0.00 &  0.00 \\
    \zro &  \zro &  \zro &  \zro & -0.001 \\
 \end{array} \right)
  \left( \begin{array}{c}
  A_0 \\ M_0 \\ M_1 \\ M_2 \\ M_3 \\
  \end{array} \right)
  \label{P4ME2}
\end{eqnarray}

\addii{For benchmark point P7 in  Class A and B (with $\tan\beta=39.7$), we have }
\begin{eqnarray}
 A_t &=& 0.31A_0 -1.09M_3 -0.03M_1 -0.11M_2 \label{P7At}\\ 
 A_\tau &=& 0.50A_0 -0.45M_2 -0.12M_1 +0.242M_3 \label{P7Atau}\\ 
 A_\mu &=& 0.65A_0 -0.50M_2 -0.13M_1 +0.257M_3 \label{P7Amu}\\ 
 M^{\rm SUSY}_1 &=&  0.46M_1  -0.007M_3 \label{P7M1}\\ 
 M^{\rm SUSY}_2 &=&  0.83M_2  -0.017M_3 \label{P7M2}\\ 
 M^{\rm SUSY}_3 &=&  1.93M_3  -0.17M_2 \label{P7M3}
\end{eqnarray}
\begin{eqnarray}
M_{\tilde{Q}_3}^2 &=&
  \left( \begin{array}{ccccc}
  A_0 & M_0 & M_1 & M_2 & M_3 \\
  \end{array} \right)
  \left( \begin{array}{ccccc}
   -0.07 &  0.02 &  0.01 & -0.01 &  0.13 \\
    \zro &  3.09 & -0.21 & -3.86 &  0.59 \\
    \zro &  \zro &  0.03 & -0.16 & -0.02 \\
    \zro &  \zro &  \zro &  4.06 & -1.21 \\
    \zro &  \zro &  \zro &  \zro &  1.96 \\
 \end{array} \right)
  \left( \begin{array}{c}
  A_0 \\ M_0 \\ M_1 \\ M_2 \\ M_3 \\
  \end{array} \right)
 \label{P7MQ3}
\end{eqnarray}
\begin{eqnarray}
M_{\tilde{U}_3}^2 &=&
  \left( \begin{array}{ccccc}
  A_0 & M_0 & M_1 & M_2 & M_3 \\
  \end{array} \right)
  \left( \begin{array}{ccccc}
   -0.06 &  0.03 &  0.01 & -0.00 &  0.11 \\
    \zro &  3.12 & -0.21 & -4.00 &  0.62 \\
    \zro &  \zro &  0.08 & -0.16 & -0.02 \\
    \zro &  \zro &  \zro &  3.81 & -1.30 \\
    \zro &  \zro &  \zro &  \zro &  1.96 \\
 \end{array} \right)
  \left( \begin{array}{c}
  A_0 \\ M_0 \\ M_1 \\ M_2 \\ M_3 \\
  \end{array} \right)
\label{P7MU3}
\end{eqnarray}
\begin{eqnarray}
M_{H_u}^2 &=&
  \left( \begin{array}{ccccc}
  A_0 & M_0 & M_1 & M_2 & M_3 \\
  \end{array} \right)
\left( \begin{array}{ccccc}
   -0.09 & -0.01 &  0.01 &  0.07 &  0.18 \\
    \zro & -1.57 &  0.17 &  2.48 & -0.37 \\
    \zro &  \zro & -0.02 &  0.10 & -0.01 \\
    \zro &  \zro &  \zro & -2.30 &  0.71 \\
    \zro &  \zro &  \zro &  \zro & -0.899 \\
 \end{array} \right)
  \left( \begin{array}{c}
  A_0 \\ M_0 \\ M_1 \\ M_2 \\ M_3 \\
  \end{array} \right)
 \label{P7MHu}
\end{eqnarray}
\begin{eqnarray}
M_{H_d}^2 &=&
  \left( \begin{array}{ccccc}
  A_0 & M_0 & M_1 & M_2 & M_3 \\
  \end{array} \right)
\left( \begin{array}{ccccc}
   -0.15 & -0.02 &  0.01 &  0.08 &  0.24 \\
    \zro & -1.46 &  0.19 &  2.50 & -0.38 \\
    \zro &  \zro & -0.01 &  0.10 &  0.00 \\
    \zro &  \zro &  \zro & -2.45 &  0.96 \\
    \zro &  \zro &  \zro &  \zro & -0.810 \\
 \end{array} \right)
  \left( \begin{array}{c}
  A_0 \\ M_0 \\ M_1 \\ M_2 \\ M_3 \\
  \end{array} \right)
 \label{P7MHd}
\end{eqnarray}
\begin{eqnarray}
\mu^2 &=&
  \frac{M^2_{H_d}-M^2_{H_u}\tan^2\beta}{\tan^2\beta-1}-\frac{m_Z^2}{2}
  \nonumber\\
  &\approx&
  \left( \begin{array}{ccccc}
  A_0 & M_0 & M_1 & M_2 & M_3 \\
  \end{array} \right)
\left( \begin{array}{ccccc}
    0.09 &  0.01 & -0.01 & -0.07 & -0.18 \\
    \zro &  1.59 & -0.17 & -2.50 &  0.37 \\
    \zro &  \zro &  0.02 & -0.10 &  0.01 \\
    \zro &  \zro &  \zro &  2.30 & -0.70 \\
    \zro &  \zro &  \zro &  \zro &  0.899 \\
 \end{array} \right)
  \left( \begin{array}{c}
  A_0 \\ M_0 \\ M_1 \\ M_2 \\ M_3 \\
  \end{array} \right)
 \label{P7mu} 
\end{eqnarray}
\begin{eqnarray}
m_{A}^2 &=&
  \frac{\tan^2\beta+1}{\tan^2\beta-1}\left(M^2_{H_d}-M^2_{H_u}\right)-m_Z^2
  \nonumber\\
  &\approx&
  \left( \begin{array}{ccccc}
  A_0 & M_0 & M_1 & M_2 & M_3 \\
  \end{array} \right)
  \left( \begin{array}{ccccc}
   -0.07 & -0.01 &  0.00 &  0.01 &  0.06 \\
    \zro &  0.11 &  0.02 &  0.01 & -0.00 \\
    \zro &  \zro &  0.01 & -0.00 &  0.01 \\
    \zro &  \zro &  \zro & -0.15 &  0.25 \\
    \zro &  \zro &  \zro &  \zro &  0.089 \\
 \end{array} \right)
  \left( \begin{array}{c}
  A_0 \\ M_0 \\ M_1 \\ M_2 \\ M_3 \\
  \end{array} \right)
 \label{P7MA}
\end{eqnarray}
\begin{eqnarray}
M_{\tilde{L}_3}^2 &=&
  \left( \begin{array}{ccccc}
  A_0 & M_0 & M_1 & M_2 & M_3 \\
  \end{array} \right)
  \left( \begin{array}{ccccc}
   -0.02 & -0.00 &  0.01 &  0.02 & -0.01 \\
    \zro &  0.83 & -0.01 & -0.01 & -0.00 \\
    \zro &  \zro &  0.03 & -0.00 &  0.00 \\
    \zro &  \zro &  \zro &  0.39 & -0.02 \\
    \zro &  \zro &  \zro &  \zro &  0.009 \\
 \end{array} \right)
  \left( \begin{array}{c}
  A_0 \\ M_0 \\ M_1 \\ M_2 \\ M_3 \\
  \end{array} \right)
 \label{P7ML3} 
\end{eqnarray}
\begin{eqnarray}
M_{\tilde{L}_2}^2 &=&
  \left( \begin{array}{ccccc}
  A_0 & M_0 & M_1 & M_2 & M_3 \\
  \end{array} \right)
  \left( \begin{array}{ccccc}
    0.00 & -0.00 & -0.00 & -0.00 & -0.00 \\
    \zro &  0.96 & -0.00 &  0.04 & -0.01 \\
    \zro &  \zro &  0.03 &  0.00 & -0.00 \\
    \zro &  \zro &  \zro &  0.35 &  0.00 \\
    \zro &  \zro &  \zro &  \zro & -0.005 \\
 \end{array} \right)
  \left( \begin{array}{c}
  A_0 \\ M_0 \\ M_1 \\ M_2 \\ M_3 \\
  \end{array} \right)
 \label{P7ML2} 
\end{eqnarray}
\begin{eqnarray}
M_{\tilde{E}_3}^2 &=&
  \left( \begin{array}{ccccc}
  A_0 & M_0 & M_1 & M_2 & M_3 \\
  \end{array} \right)
  \left( \begin{array}{ccccc}
   -0.05 & -0.00 &  0.01 &  0.04 & -0.02 \\
    \zro &  0.74 & -0.01 & -0.10 &  0.01 \\
    \zro &  \zro &  0.12 & -0.01 &  0.00 \\
    \zro &  \zro &  \zro &  0.06 & -0.04 \\
    \zro &  \zro &  \zro &  \zro &  0.026 \\
 \end{array} \right)
  \left( \begin{array}{c}
  A_0 \\ M_0 \\ M_1 \\ M_2 \\ M_3 \\
  \end{array} \right)
 \label{P7ME3} 
\end{eqnarray}
\begin{eqnarray}
M_{\tilde{E}_2}^2 &=&
  \left( \begin{array}{ccccc}
  A_0 & M_0 & M_1 & M_2 & M_3 \\
  \end{array} \right)
  \left( \begin{array}{ccccc}
   -0.00 & -0.00 & -0.00 & -0.00 & -0.00 \\
    \zro &  0.99 & -0.00 &  0.01 & -0.00 \\
    \zro &  \zro &  0.13 &  0.00 & -0.00 \\
    \zro &  \zro &  \zro & -0.00 &  0.00 \\
    \zro &  \zro &  \zro &  \zro & -0.001 \\
 \end{array} \right)
  \left( \begin{array}{c}
  A_0 \\ M_0 \\ M_1 \\ M_2 \\ M_3 \\
  \end{array} \right)
  \label{P7ME2}
\end{eqnarray}
\end{spacing}

\section*{Appendix B: The detail information of the 7 benchmark points}
\vspace*{-0.3cm}
As a supplement to Tab.\ref{table2}, we list in Tab.\ref{table3} the model parameters, relevant sparticle masses and phenomenological observables for the 7 benchmark points.
\vspace*{-0.3cm}
\begin{table}[!htb]
\centering
\caption{The detail information of the 7 benchmark points. }
\label{table3}
\begin{spacing}{1.2}
\begin{tabular}{|c|c|c|c|c|c|c|c|}
\hline
    & ~~~~P1~~~~ & ~~~~P2~~~~ & ~~~~P3~~~~  & ~~~~P4~~~~  & ~~~~P5~~~~ & ~~~~P6~~~~  & ~~~~P7~~~~  \\ \hline
 $\tan\beta$ & 54.1 & 28.5 & 29.1 & 20.8 & 58.3 & 48.5 & 39.7
 \\ \hline
 $A_0$ [GeV] & 2607 & 2410 & -72 & -2250 & 1516 & -1227 & 363
 \\ \hline
 $M_0$ [GeV] & 979 & 808 & 736 & 564 & 718 & 624 & 577
 \\ \hline
 $M_1$ [GeV] & 772 & 850 & 813 & 291 & 548 & 773 & 712
 \\ \hline
 $M_2$ [GeV] & -819 & 376 & -685 & 971 & -988 & 967 & -713
 \\ \hline
 $M_3$ [GeV] & 9881 & -6080 & -8350 & 6335 & 6032 & 6734 & -9137
 \\ \hline
 \hline
 $\mu$ [GeV] & 9075 & 6338 & 8081 & 6619 & 5788 & 6808 & 8786 \\ \hline
 $m_h$ [GeV] & 126.2 & 124.9 & 125.2 & 125.1 & 125.2 & 125.8 & 125.7 \\ \hline
 $m_A$ [GeV] & 6861 & 5011 & 6570 & 6407 & 4276 & 5145 & 2596 \\ \hline
 $m_{\tilde{t}_1}$ [GeV] & 13819 & 8634 & 11742 & 8870 & 8720 & 9491 & 12744 \\ \hline
 $m_{\tilde{\tau}_1}$ [GeV] & 336 & 543 & 477 & 120 & 250 & 350 & 1015 \\ \hline
 $m_{\tilde{\mu}_1}$ [GeV] & 912 & 754 & 627 & 540 & 707 & 641 & 388 \\ \hline
 $m_{\tilde{\nu}_1}$ [GeV] & 926 & 751 & 625 & 672 & 874 & 718 & 390 \\ \hline
 $m_{\tilde{\chi}^\pm_1}$ [GeV] & 920 & 451 & 460 & 738 & 986 & 734 & 485 \\ \hline
 $Br(B_s\to\mu^+\mu^-) [10^{-9}]$ & 3.67 & 3.53 & 3.55 & 3.58 & 3.89 & 3.75 & 3.20
 \\ \hline
 $Br(B_d\to\mu^+\mu^-) [10^{-10}]$ & 1.05 & 1.00 & 1.00 & 1.01 & 1.14 & 1.06 & 0.87
 \\ \hline
 $Br(b\to s\gamma) [10^{-4}]$ & 3.35 & 3.38 & 3.37 & 3.36 & 3.36 & 3.36 & 3.44
 \\ \hline
 $\Delta a_{\mu}^{\rm SUSY} [10^{-10}]$ & 7.81 & 6.51 & 10.3 & 6.36 & 7.95 & 18.0 & 46.0
 \\ \hline
 $m_{\tilde{\chi}^0_1}$ [GeV] & 284.8  & 423.8  & 426.9  & 86.9  & 205.2  & 303.8  & 386.0
 \\ \hline
 $m_{\tilde{\chi}^0_2}$ [GeV] & 920  & 451  & 460  & 738  & 986  & 734  & 485
 \\ \hline
 $\sigma_{\rm SI}~[10^{-50}\;{\rm cm}^{-1}]$
 & 0.91 & 6.63 & 3.04 & 3.58 & 3.16 & 2.68 & 2.23
 \\ \hline
 $\sigma^{\rm SD}_{\rm P}~[10^{-47}\;{\rm cm}^{-1}]$
 & 1.29 & 5.31 & 1.99 & 4.17 & 7.82 & 3.95 & 1.44
 \\ \hline
 $\sigma^{\rm SD}_{\rm N}~[10^{-47}\;{\rm cm}^{-1}]$
 & 0.97 & 3.47 & 1.40 & 2.62 & 5.81 & 2.65 & 1.01
 \\ \hline
 $\Omega h^2$ & 0.109  & 0.112  & 0.120  & 0.111  & 0.130  & 0.126  & 0.118 \\ \hline
\end{tabular}
\end{spacing}
\end{table}

\end{document}